\documentclass[twocolumn,amsmath,nofootinbib]{revtex4} 

\usepackage{url}
\usepackage{amssymb}
\usepackage{amsfonts}
\usepackage{amsbsy}
\usepackage{graphicx}
\usepackage{hyperref}
\usepackage{array} 
\usepackage{multirow}

\newcolumntype{x}[1]{%
>{\centering\hspace{0pt}}p{#1}}%
\providecommand{\openone}{\leavevmode\hbox{\small1\kern-3.8pt\normalsize1}}

\def\etal{{\frenchspacing\it et al.}}
\def\ie{{\frenchspacing\it i.e.}}
\def\eg{{\frenchspacing\it e.g.}}
\def\etc{{\frenchspacing\it etc.}}

\def\spose#1{\hbox to 0pt{#1\hss}}
\def\simlt{\mathrel{\spose{\lower 3pt\hbox{$\mathchar"218$}}
   \raise 2.0pt\hbox{$\mathchar"13C$}}}
\def\simgt{\mathrel{\spose{\lower 3pt\hbox{$\mathchar"218$}}
     \raise 2.0pt\hbox{$\mathchar"13E$}}}
 \def\simpropto{\mathrel{\spose{\lower 3pt\hbox{$\mathchar"218$}}
     \raise 2.0pt\hbox{$\propto$}}}

\def\beq#1{\begin{equation}\label{#1}}
\def\eeq{\end{equation}}
\def\beqa#1{\begin{eqnarray}\label{#1}}
\def\eeqa{\end{eqnarray}}
\def\eq#1{equation~(\ref{#1})}	
\def\Eq#1{Equation~(\ref{#1})}
\def\eqn#1{~(\ref{#1})}

\def\fig#1{Figure~\ref{#1}}
\def\Fig#1{Figure~\ref{#1}}

\def\Sec#1{Section~\ref{#1}}
\def\Sec#1{Section~\ref{#1}}

\def\ed{\end{document}}

\def\astar{a_*}
\def\adeath{a_\dagger}
\def\alimit{a_{\rm GRB}}

\def\E{{\bf E}}

\def\P{{P}}
\def\PP{\Pi}
\def\U{{U}}

\def\rplanck{r_{\rm pl}}

\def\Hsubj{H_{\rm s}}
\def\Hobj{H_{\rm o}}
\def\Henv{H_{\rm e}}
\def\Hso{H_{\rm so}}
\def\Hoe{H_{\rm oe}}
\def\Hse{H_{\rm se}}
\def\Hsoe{H_{\rm soe}}
\def\Sobj{S_{\rm o}}

\def\robj{\rho_{\rm o}}

\def\rso{\rho_{\rm so}}

\def\tr{\hbox{tr}\,}

\def\tensormult{\otimes}
\def\expec#1{\langle#1\rangle}

\def\blankface{\,\ddot{\raisebox{-2pt}{-}}\,}
\def\smileface{\raisebox{-2pt}{$\ddot\smile$}}
\def\frownface{\raisebox{-2pt}{$\ddot\frown$}}

\def\ket#1{|#1\rangle}
\def\psiket{\ket{\psi}}
\def\noobs{\ket{\blankface}}
\def\upobs{\ket{\smileface}}
\def\downobs{\ket{\frownface}}

\def\up{\ket{\!\!\uparrow}}
\def\down{\ket{\!\!\downarrow}}

\def\noup{\ket{\blankface\!\!\uparrow}}
\def\nodown{\ket{\blankface\!\!\downarrow}}
\def\upup{\ket{\smileface\!\!\uparrow}}
\def\updown{\ket{\smileface\!\!\downarrow}}
\def\downup{\ket{\frownface\!\!\uparrow}}
\def\downdown{\ket{\frownface\!\!\downarrow}}

\def\bra#1{\langle #1|}

\def\noobsbra{\bra{\blankface}}

\def\upbra{\bra{\uparrow\!\!}}
\def\downbra{\bra{\downarrow\!\!}}

\def\noupbra{\bra{\blankface\!\!\uparrow\!\!}}
\def\nodownbra{\bra{\blankface\!\!\downarrow\!\!}}
\def\upupbra{\bra{\smileface\!\!\uparrow\!\!}}

\def\downdownbra{\bra{\frownface\!\!\downarrow}}

\def\rhohappy{\rho_{   \bigcirc\hskip-3.8mm{   \raisebox{2.5pt}{  \tiny{\raisebox{-2pt}{$\ddot\smile$}   }   }}   }}
\def\rhosad{\rho_{   \bigcirc\hskip-3.8mm{   \raisebox{2.5pt}{  \tiny{\raisebox{-2pt}{$\ddot\frown$}   }   }}   }}

\def\rn{}
\def\nn#1 #2{#2. #1}				
\def\nnn#1 #2 #3{#2. #3. #1}			
\def\nnnn#1 #2 #3 #4{#2. #3. #4 #1}		
\def\nnnnn#1 #2 #3 #4 #5{#2. #3. #4 #5. #1}	
\def\dualand{ and\hbox{ }}				
\def\multiand{, and\hbox{ }}				
\def\rf#1;#2;#3;#4;#5 {{\frenchspacing\par\rn#1, #3 {\bf #4}, #5 (#2). \par}}
\def\rg#1;#2;#3;#4;#5;#6 {{\frenchspacing\par\rn#1, #3 {\bf #4}, #5 (#2). \par}}
\def\rfbook#1;#2;#3;#4;#5 {{\frenchspacing\par\rn#1, {\it #3} (#5, #4, #2).\par}}
\def\rfprep#1;#2;#3 {{\par\frenchspacing\rn#1, #3 (#2).\par}}
\def\rfproc#1;#2;#3;#4;#5;#6 {{\frenchspacing\par\rn#1 #2, in {\it #3}, ed. #4 (#5: #6)\par}}
\def\rfprocp#1;#2;#3;#4;#5;#6;#7 {{\frenchspacing\par\rn#1 #2, in {\it #3}, ed. #4 (#5: #6), p#7\par}}

\begin{document}
\pdfoptionalwaysusepdfpagebox=5


\title{How unitary cosmology generalizes thermodynamics and\\solves the inflationary entropy problem}

\author{Max Tegmark}

\address{Dept.~of Physics \& MIT Kavli Institute, Massachusetts Institute of Technology, Cambridge, MA 02139}

\date{Physical Review D {\bf 85}, 123517, submitted August 27 2011, accepted March 20 2012, published June 11}

\vspace{10mm}

\begin{abstract}
We analyze cosmology assuming unitary quantum mechanics, using a tripartite partition into system, observer and environment degrees of freedom.
This generalizes the second law of thermodynamics to {\it ``The system's entropy can't decrease unless it interacts with the observer, and it can't increase unless it interacts with the environment.''}
The former follows from the quantum Bayes Theorem we derive.
We show that because of the long-range entanglement created by cosmological inflation, the cosmic entropy decreases exponentially rather than linearly with the number of bits of information observed, so that a given observer can reduce entropy by much more than the amount of information her brain can store. Indeed, we argue that as long as inflation has occurred in a non-negligible fraction of the volume, almost all sentient observers will find themselves in a post-inflationary low-entropy Hubble volume, and we humans have no reason to be surprised that we do so as well, which solves the so-called inflationary entropy problem.
An arguably worse problem for unitary cosmology involves gamma-ray-burst constraints on the ``Big Snap'', a fourth cosmic doomsday scenario alongside the ``Big Crunch'', ``Big Chill'' and ``Big Rip'', where an increasingly granular nature of expanding space modifies our life-supporting laws of physics.
    Our tripartite framework also  clarifies when the popular quantum gravity approximation $G_{\mu\nu}\approx 8\pi G \expec{T_{\mu\nu}}$ is valid, and how problems with recent attempts to explain dark energy as gravitational backreaction from super-horizon scale fluctuations can be understood as a failure of this approximation.
\end{abstract}

\maketitle

\section{Introduction}

The spectacular progress in observational cosmology over the past decade has established 
cosmological inflation \cite{Guth81,Linde82,AlbrechtSteinhardt82,Linde83} as the most popular theory for what happened early on.
Its popularity stems from the perception that it elegantly explains certain observed properties of our universe that would otherwise constitute extremely unlikely fluke coincidences, such as why it is so flat and uniform,  and why there are $10^{-5}$-level density fluctuations which appear adiabatic, Gaussian, and almost scale-invariant \cite{Spergel03,sdsspars,wmap7}.

\begin{figure}[pbt]
\centerline{\includegraphics[width=92mm]{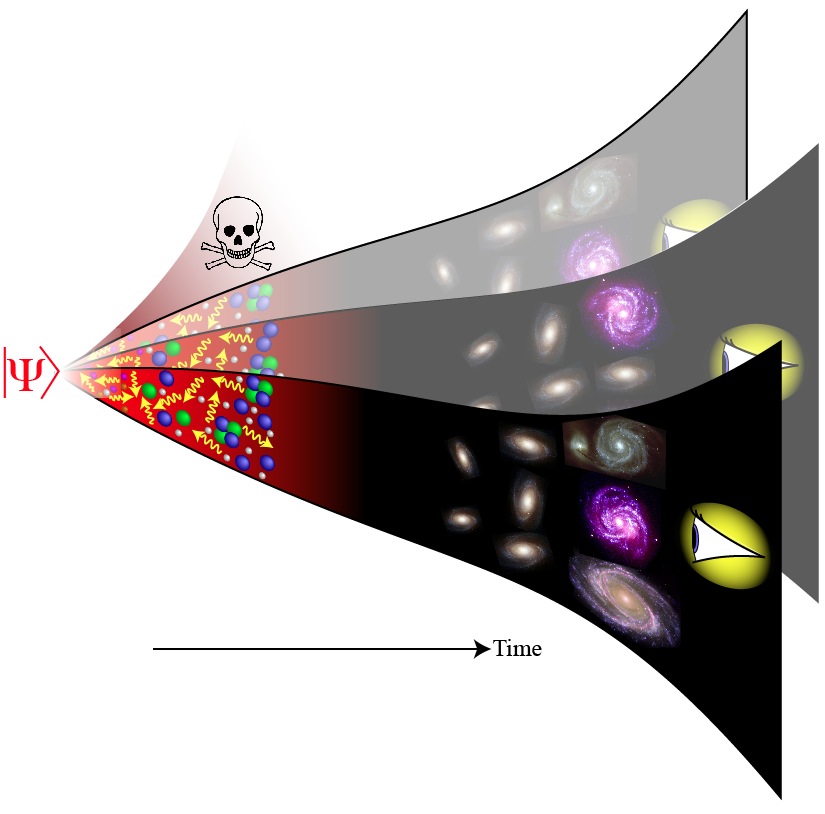}}
\vskip-3mm
\caption{Because of chaotic dynamics, a single early-universe quantum state $\psiket$
typically evolves into a quantum superposition of many macroscopically different states, some of which correspond to a large semiclassical post-inflationary universe like ours (each with its galaxies {\protect\etc} in different places), and others which do not and completely lack observers.
}
\label{ExpandingUniverseSplitFig}
\end{figure}

If a scientific theory predicts a certain outcome with probability below $10^{-6}$, say, then  we say that the theory is ruled out at 99.9999\% confidence if we nonetheless observe this outcome. In this sense, the classic Big Bang model without inflation is arguably ruled out at extremely high significance. For example, generic initial conditions consistent with our existence 13.7 Billion years later predict observed cosmic background fluctuations that are about $10^5$ times larger than we actually observe \cite{inflation} --- the so-called horizon problem \cite{Guth81}. In other words, without inflation, the initial conditions would have to be highly fine-tuned to match our observations.

However, the case for inflation is not yet closed, even aside from issues to do with measurements \cite{Copi10}, 
competing theories \cite{PreBigBang93, Ekpyrotic01,BrandenbergerStringGas09} 
and the so-called measure problem 
\cite{Linde95,multiverse4wheeler,conditionalization,inflation,Freivogel05,Garriga05,Easther05,axions,Aguirre06a, Aguirre06b,Bousso06,GibbonsTurok08,HartleHertog07,Page08,deSimone08,Linde09,Bousso10,Linde10, Page10,Vilenkin11,Salem11,GuthVanchurin11}.
 In particular, it has been argued that the so-called ``entropy problem'' invalidates claims that inflation is a successful theory. 
This ``entropy problem'' was articulated by Penrose even before inflation was invented \cite{Penrose79}, and has recently been clarified in an important body of work by Carroll and collaborators \cite{CarrollBook,CarrollTam10}.
The basic problem is to explain why our early universe had such low entropy, with its matter highly uniform rather than clumped into huge black holes.
The conventional answer holds that inflation is an attractor solution, such that a broad class of initial conditions lead to essentially the same inflationary outcome, thus replacing the embarrassing need to explain extremely unusual initial conditions by the less embarrassing need to explain why our initial conditions were in the broad class supporting inflation. 
However, \cite{CarrollTam10} argues that the entropy must have been at least as low {\it before} inflation as after it ended, so that inflation fails to make our state seem less unnatural or fine-tuned. 
This follows from the mapping between initial states and final states being invertible, corresponding to Liouville's theorem in classical mechanics and unitarity in quantum mechanics. 

The main goal of this paper is to investigate the entropy problem in unitary quantum mechanics more thoroughly.
We will see that this fundamentally transforms the problem, strengthening the case for inflation.
A secondary goal is to explore other implications of unitary cosmology, for example
by clarifying when the popular approximation $G_{\mu\nu}\approx 8\pi G \expec{T_{\mu\nu}}$ is and is not valid. 
The rest of this paper is organized as follows. 
In  \Sec{TripartiteSec}, we describe a quantitative formalism for computing the quantum state 
and its entropy in unitary cosmology. 
We apply this formalism to the inflationary entropy problem in \Sec{EntropyProbSec} and discuss implications in \Sec{DiscussionSec}.
Details regarding the ``Big Snap'' scenario are covered to Appendix~\ref{SnapSec}.

\section{Subject, object \& environment}
\label{TripartiteSec}

\subsection{Unitary Cosmology}

The key assumption underlying the entropy problem is that quantum mechanics is unitary, so we will make this assumption throughout the present paper\footnote{The forms of non-unitarity historically invoked to address the quantum measurement problem tend to make the entropy problem worse rather than better: both Copenhagen-style wavefunction collapse \cite{Bohr,Heisenberg} and proposed dynamical reduction mechanisms \cite{GRW86} arguably tend to {\it increase} the entropy, transforming pure (zero entropy) quantum states into mixed states, akin to a form of diffusion process in phase space.}.
As described in \cite{nihilo}, this suggests the history schematically illustrated in \fig{ExpandingUniverseSplitFig}:
a wavefunction describing an early universe quantum state (illustrated by the fuzz at the far left)
will evolve deterministically according to the Schr\"odinger equation into a quantum superposition of not one but many 
macroscopically different states, some of which correspond to large semiclassical post-inflationary universes like ours, and others which do not and completely lack observers.
The argument of \cite{nihilo} basically went as follows:
\begin{enumerate}
\item
By the Heisenberg uncertainty principle, any initial state must involve micro-superpositions, microscopic quantum 
fluctuations in the various fields.
\item
Because the  ensuing time-evolution involves instabilities (such as the well-known gravitational
instabilities that lead to the formation of cosmic large-scale structure), some of these
micro-superpositions are amplified into macro-superpositions, much like in Schr\"odinger's cat example \cite{Schrodinger}. More generally, this happens for any chaotic dynamics, where positive Lyapunov exponents make the outcome exponentially sensitive to initial conditions. 
\item
The current quantum state of the universe is thus a superposition of a large number of states that
are macroscopically different (Earth forms here, Earth forms one meter further north,  {\it etc}), as well as states that failed to inflate.
\item
Since macroscopic objects inevitably interact with their surroundings,
the well-known effects of decoherence will keep 
observers such as us unaware of such macro-superpositions.
\end{enumerate}
This shows that with unitary quantum mechanics, the conventional phrasing of the entropy problem is too simplistic, 
since a single pre-inflationary quantum state evolves into a superposition of many different semiclassical post-inflationary states.
The careful and detailed analysis of the entropy problem in \cite{CarrollTam10} is mainly performed within the context of classical physics, and quantum mechanics is only briefly mentioned, when correctly stating that Liouville's theorem holds quantum mechanically too as long as the evolution is unitary. However, the evolution that is unitary is that of the {\it total} quantum state of the entire universe.
We unfortunately have no observational information about this total entropy, and what we casually refer to as ``the'' entropy is instead the entropy we observe for our particular branch of the wavefunction in \fig{ExpandingUniverseSplitFig}.
We should generally expect these two entropies to be quite different --- indeed, the entropy of the entire universe may well equal zero, since if it started in a pure state, unitarity ensures that it is still in a pure state.

\subsection{Deconstructing the universe}

It is therefore interesting to investigate the cosmological entropy problem more thoroughly in the context of unitary quantum mechanics, which we will now do. 

\begin{figure}[pbt]
\centerline{\includegraphics[width=88mm]{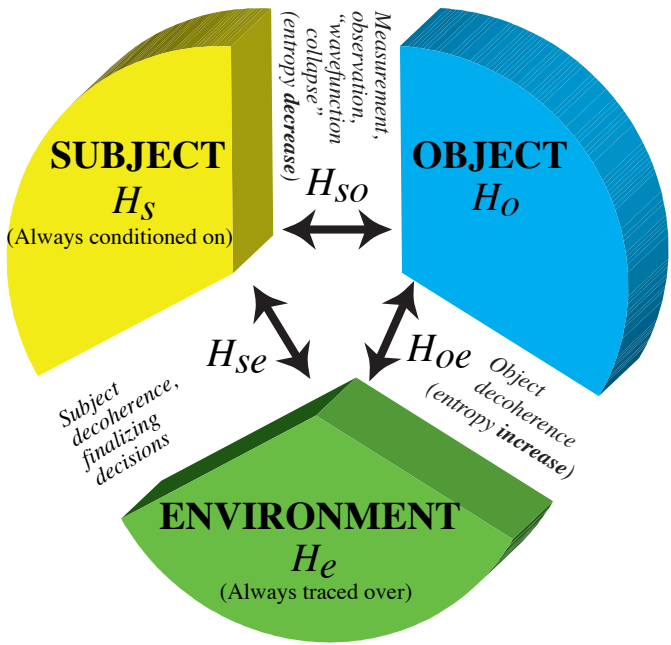}}
\caption{
An observer 
can always decompose the world into three subsystems:
the degrees of freedom corresponding to her subjective
perceptions (the subject),
the degrees of freedom being studied (the object), 
and everything else (the environment).
As indicated, the subsystem Hamiltonians 
$\Hsubj$, $\Hobj$, $\Henv$ and 
the interaction Hamiltonians
$\Hso$, $\Hoe$, $\Hse$ can
cause qualitatively very different effects,
providing a unified picture including both decoherence and 
apparent wave function collapse.
Generally, $\Hoe$ increases entropy and $\Hso$ decreases entropy. 
}
\label{TrinityFig}
\end{figure}

Most discussions of quantum statistical mechanics
split the Universe into two subsystems \cite{Feynman72}:
the object under consideration and everything else
(referred to as the {\it environment}).
At a physical level, this ``splitting'' is simply a matter of accounting, grouping the degrees of freedom into two sets: those of the object and the rest. At a mathematical level, this corresponds to a choice of factorization of the Hilbert space.

As discussed in \cite{brain}, unitary quantum mechanics can be even better understood if we include a third subsystem as well, the {\it subject}, thus decomposing the total system (the entire universe) into  {\it three}  subsystems, as illustrated in \fig{TrinityFig}: 
\begin{enumerate}
\item The {\bf subject} consists of the degrees 
of freedom associated with the 
subjective perceptions of the observer.
This does not include any other degrees of freedom associated
with the brain or other parts of the body.
\item The {\bf object} consists of the degrees of freedom 
that the observer is interested in studying, \eg, the pointer
position on a measurement apparatus.
\item The {\bf environment} consists of everything else,
\ie, all the degrees of freedom that the observer is 
not paying attention to.
By definition, these are the degrees of freedom that we 
always perform a partial trace over.
\end{enumerate}

A related framework is presented in  \cite{brain,Nomura11}.
Note that the first two definitions are very restrictive.
Suppose, for example, that you are measuring a voltage using one of those old-fashioned multimeters that has an analog pointer.
Then the ``object''  consists merely of the single degree of freedom corresponding to the angle of the pointer, and excludes all of the
other $\sim 10^{27}$ degrees of freedom associated with the atoms in the multimeter.
Similarly, the ``subject''  excludes most of the  $\sim 10^{28}$ degrees of freedom associated with the elementary particles in your brain.
The term ``perception'' is used in a broad sense in item~1, 
including thoughts, emotions and any other attributes of the 
subjectively perceived state of the observer.

\def\zupstate{
\setlength{\extrarowheight}{2pt}
\left(
\begin{tabular}{cc}
$1$	&$0$\\
$0$	&$0$
\end{tabular}
\right)
}

\def\xupstate{
\setlength{\extrarowheight}{2pt}
\left(
\begin{tabular}{cc}
${1\over 2}$	&${1\over 2}$\\
${1\over 2}$	&${1\over 2}$
\end{tabular}
\right)
}

\def\mixedstate{
\setlength{\extrarowheight}{2pt}
\left(
\begin{tabular}{cc}
${1\over 2}$	&$0$\\
$0$		&${1\over 2}$
\end{tabular}
\right)
}

\begin{table*}
\caption{Summary of three three basic quantum processes discussed in the text}
\setlength{\extrarowheight}{13pt}
\begin{tabular}{|l|l|c|l|l|}
\hline
Interaction		&Dynamics							&Example						&Effect			&Entropy\\
\hline
Object-object		&$\rho\mapsto\U\rho\U^\dagger$			&$\zupstate\mapsto\xupstate$		&Unitary evolution	&Unchanged\\
Object-environment	&$\rho\mapsto\sum_{ij}\P_i\rho\P_j\langle\epsilon_j|\epsilon_i\rangle$			&$\xupstate\mapsto\mixedstate$	&Decoherence		&Increases\\
Object-subject		&$\rho\mapsto {\PP_i\rho\PP^\dagger_i\over\tr\PP_i\rho\PP_i^\dagger}, \quad \PP_i=\sum_j  \langle s_i|\sigma_j\rangle\P_j$	&$\mixedstate\mapsto\zupstate$	&Observation		&Decreases\\
\hline
\end{tabular}
\end{table*}

Just as with the currently standard bipartite decomposition into object and environment, this tripartite decomposition is different for each observer and situation: the subject degrees of freedom depend on which of the many observers in our universe is doing the observing, the object degrees of freedom reflect which physical system this observer chooses to study, and the environment degrees of freedom correspond to everything else. For example, if you are studying an electron double-slit experiment, electron positions would constitute your object and decoherence-inducing photons would be part of your environment, whereas in many quantum optics experiments, photon degrees of freedom are the object while electrons are part of the environment.

This  subject-object-environment decomposition of the degrees of freedom allows a corresponding 
decomposition of the Hamiltonian:
\beq{HdecompEq}
H = H_s + H_o + H_e + \Hso + \Hse + \Hoe + \Hsoe,
\eeq
where the first three terms operate only within one subsystem, the second three terms represent pairwise interactions between subsystems, and the third term represents any irreducible three-way interaction.
The practical usefulness of this tripartite
decomposition lies in that 
one can often neglect everything except the object
and its internal dynamics (given by $\Hobj$) to first order,
using simple prescriptions to correct for the interactions
with the subject and the environment, as summarized in Table~1.
The effects of both $\Hso$ and
$\Hoe$ have been extensively studied in the literature.
$\Hso$ involves quantum measurement, 
and gives rise to the usual interpretation of the diagonal elements
of the object density matrix as probabilities.
$\Hoe$ produces decoherence, selecting a preferred basis
and making the object act classically under appropriate conditions.
$\Hse$, causes decoherence directly in the subject system.
For example, \cite{brain} showed that any qualia or other subjective perceptions that 
are related to neurons firing in a human brain will decohere extremely rapidly, typically on a timescale of order $10^{-20}$ seconds, 
ensuring that our subjective perceptions will appear classical. 
In other words, it is useful to split the Schr\"odinger equation into pieces: three governing the three parts of our universe (subject, object and environment), and additional pieces governing the interactions between these parts.
Analyzing the effects of these different parts of the equation, 
the $\Hobj$ part gives most of the effects that our textbooks cover, 
the $\Hso$ part gives Everett's many worlds (spreading superpositions from the object to you, the subject), 
the $\Hoe$ part gives traditional decoherence, 
the $\Hse$ part gives subject decoherence.

\subsection{Entropy in quantum cosmology}

In the context of unitary cosmology, this tripartite decomposition is useful not merely as a framework for classifying and unifying different quantum effects, but also as a framework for understanding entropy and its evolution.
In short, $\Hoe$ increases entropy while $\Hso$ decreases entropy, in the sense defined below.

To avoid confusion, it is crucial that we never talk of  {\it the} entropy without being clear on which entropy we are referring to.
With three subsystems, there are many interesting entropies to discuss, for example that of the subject, that of the object, that of the environment and that of the whole system, all of which will generally be different from one another. 
Any given observer can describe the state of an object of interest by a density matrix $\robj$ which is computed from the 
full density matrix $\rho$ in two steps:
\begin{enumerate}
\item {\bf Tracing:} Partially trace over all environment degrees of freedom.
\item{\bf Conditioning:} Condition on all subject degrees of freedom.
\end{enumerate}
In practice, step 2  often reduces to what textbooks call ``state preparation'', as explained below.
When we say ``the entropy'' without further qualification, we will refer to the object entropy $\Sobj$: the standard von Neumann entropy of this object density matrix $\robj$, \ie, 
\beq{SdefEq}
\Sobj\equiv-\tr\robj \log\robj.
\eeq
Throughout this paper, we use logarithms of base two so that the entropy has units of bits.
Below when we speak of the information (in bits) that one system (say the environment) has about another (say the object), we 
will refer to the quantum mutual information given by the standard definition \cite{Everett57,EverettBook,NielsenChuangBook}
\beq{IdefEq}
I_{12} \equiv S_1 + S_2 - S_{12},
\eeq
where $S_{12}$ is the joint system, while $S_1$ and $S_1$ are the entropies of each subsystem when tracing over the degrees of freedom of the other.

\begin{figure}[b]
\centerline{\includegraphics[width=87mm]{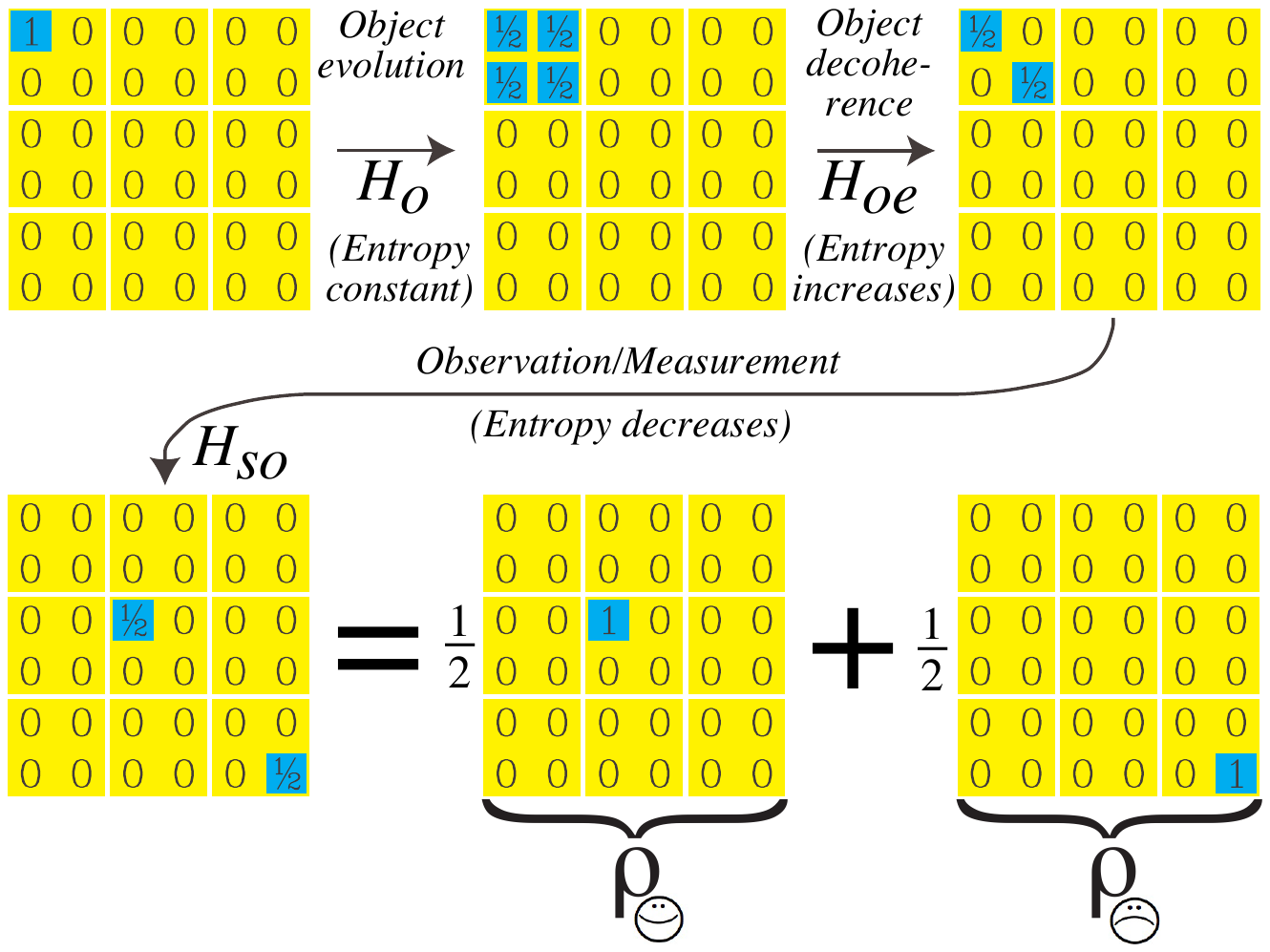}}
\bigskip
\caption{Time evolution of the $6\times6$ density matrix for the basis states 
$\protect\noup$, $\protect\nodown$,
$\protect\upup$, $\protect\updown$,
$\protect\downup$, $\protect\downdown$ 
as the object evolves in isolation, then decoheres, then gets observed by the subject. The final result is a statistical mixture of the states $\protect\upup$ and $\protect\downdown$, simple zero-entropy states like the one we started with.
}
\label{ChessFig}
\end{figure}

Let us illustrate all this with a simple example in \fig{ChessFig},
where both the subject and object
have only a single degree of freedom that can take on only a few
distinct values (3 for the subject, 2 for the object).
For definiteness, 
we denote the three subject states 
$\noobs$, $\upobs$ and $\downobs$,
and interpret them as the observer
feeling neutral, happy and sad, respectively.
We denote the two object states $\up$ and $\down$,
and interpret them as 
the spin component (``up'' or ``down'') 
in the $z$-direction of a spin-1/2 system,
say a silver atom.
The joint system consisting of subject and object therefore has
only $2\times 3=6$ basis states:
$\noup$, $\nodown$,
$\upup$, $\updown$,
$\downup$, $\downdown$.
In Figures~\fig{ChessFig}, 
we have therefore plotted $\rho$
as a $6\times 6$ matrix consisting of nine two-by-two blocks.

\subsubsection{Effect of $\Hobj$: constant entropy}

If the object were to evolve during a time interval $t$
without interacting with
the subject or the environment ($\Hso=\Hoe=\Hsoe=0$), 
then its reduced density matrix $\robj$ would evolve
into $\U\robj\U^\dagger$ with the same entropy,
since the time-evolution operator 
$\U\equiv e^{-i\Hobj t}$ is unitary.

Suppose the subject stays in the state $\noobs$ and the object
starts out in the pure state $\up$.
Let the object Hamiltonian
$\Hobj$ correspond to a magnetic field in the $y$-direction
causing the spin to precess to the $x$-direction, \ie, to the state
$(\up+\down)/\sqrt{2}$.
The object density matrix $\robj$ then evolves into
\beqa{ConstantEntropyEq}
\robj 
&=&U\up\upbra\U^\dagger = {1\over 2}(\up+\down)(\upbra+\downbra)\nonumber\\
&=&{1\over 2}(\up\upbra + \up\downbra + \down\upbra + \down\downbra),
\eeqa
corresponding to the four entries of $1/2$ in 
the second matrix of \fig{ChessFig}.

This is quite typical of pure quantum evolution:
a basis state eventually evolves into a superposition
of basis states, 
and the quantum nature of this superposition
is manifested by off-diagonal elements in $\robj$. 
Another familiar example of this is 
the familiar spreading out of the wave packet of 
a free particle.

\subsubsection{Effect of $\Hoe$: increasing entropy}

This was the effect of $\Hobj$ alone. 
In contrast, $\Hoe$ will generally cause decoherence and increase 
the entropy of the object. Although decoherence is now well-understood \cite{Zeh70,JZ85,ZehBook,Zurek09,SchlosshauerBook}, we will briefly review some core results here that will be needed for the subsequent section about measurement.

Let $\ket{o_i}$ and $\ket{e_i}$ denote basis states of the object and the environment, respectively.
As discussed in detail in \cite{ZehBook,SchlosshauerBook}, decoherence (due to $\Hoe$) tends to occur on timescales much faster than those on which macroscopic objects evolve (due to $\Hobj$), making it a good approximation to assume that
the unitary dynamics is $\U\equiv e^{-i\Hoe t}$ on the decoherence timescale and leaves the object state unchanged, merely changing the 
environment state in a way that depends on the object state $\ket{o_i}$, say from an initial state $\ket{e_0}$ into some final state $\ket{\epsilon_i}$:
\beq{DecoAssumptionEq}
U\ket{e_0}\ket{o_i}=\ket{\epsilon_i}\ket{o_i}.
\eeq
This means that the initial density matrix $\rho=\ket{e_0}\bra{e_0}\tensormult\rho_o$ of the object-environment system, where
$\rho_o=\sum_{ij}\bra{o_i}\rho_o\ket{o_j}\ket{o_i}\bra{o_j}$,
will evolve as 
\beqa{DecoDerivationEq1}
\rho&\mapsto&U\rho\U^\dagger = U\ket{e_0}\bra{e_0}\rho_o\U^\dagger\nonumber\\
       &=&\sum_{ij}\bra{o_i}\rho_o\ket{o_j}U\ket{e_0}\ket{o_i}\bra{e_0}\bra{o_j}\U^\dagger\nonumber\\
       &=&\sum_{ij}\bra{o_i}\rho_o\ket{o_j}\ket{\epsilon_i}\ket{o_i}\bra{\epsilon_j}\bra{o_j}.
\eeqa
The reduced density matrix for the object is this object-environment density matrix partial-traced over the environment, so it evolves as 
\beqa{DecoDerivationEq2}
\rho_0&\mapsto&\tr_e\rho\equiv\sum_k\bra{e_k}\rho\ket{e_k}\nonumber\\
            &=&\sum_{ijk}\bra{o_i}\rho_o\ket{o_j} \bra{\epsilon_j}e_k\rangle \bra{e_k}\epsilon_i\rangle  \ket{o_i}\bra{o_j}\nonumber\\
           &=&\sum_{ij}  \ket{o_i}\bra{o_i}\rho_o\ket{o_j}\bra{o_j} \bra{\epsilon_j}\epsilon_i\rangle\nonumber\\      
           &=&\sum_{ij} P_i \rho_o P_j\bra{\epsilon_j}\epsilon_i\rangle,            
\eeqa
where we used the identity $\sum_k \ket{e_k}\bra{e_k}=I$ in the penultimate step and defined the projection operators 
$P_i\equiv \ket{o_i}\bra{o_i}$ that project onto the $i^{\rm th}$ eigenstate of the object. 
This well-known result implies that if the environment can tell whether the object is in state $i$ or $j$, \ie, if the environment reacts differently in these two cases by ending up in two orthogonal states, $\bra{\epsilon_j}\epsilon_i\rangle=0$, then the corresponding $(i,j)$-element of the object density matrix gets replaced by zero:
\beq{vonNeumannReductionEq}
\rho_0\mapsto\sum_i P_i \rho_o P_i,            
\eeq
corresponding to the so-called von Neumann reduction \cite{VonNeumann32} which was postulated long before the discovery 
of decoherence; we can interpret it as object having been measured by something (the environment) that refuses to tell us what the outcome was.\footnote{\Eq{vonNeumannEq} is known as the L{\"u}ders projection \cite{Lueders1950} for the more general case where 
the $P_i$ are more general projection operators that still satisfy $P_i P_j=\delta_{ij}P_i$, $\sum P_i=I$.
This form also follows from the decoherence formula~\eqn{DecoDerivationEq2} for the more general case where the environment can only 
tell which group of states the object is in (because the eigenvalues of $H_{oe}$ are degenerate within each group), so that $\bra{\epsilon_j}\epsilon_i\rangle=1$ if $i$ and $j$ belong to the same group and vanishes otherwise. One then obtains an equation of the same form as \eq{vonNeumannReductionEq}, but where each projection operator projects onto one of the groups.
}

This suppression of the off-diagonal elements of the object density matrix is illustrated
in \fig{ChessFig}. 
In this example, we have only two object states $\ket{o_1}=\up$ and $\ket{o_2}=\down$, two environment states, and an interaction such 
that $\bra{\epsilon_1}\epsilon_2\rangle=0$, giving
\beq{DecoherenceExampleEq}
\rho_o\mapsto {1\over 2}(\up\upbra + \down\downbra.
\eeq
This new final state corresponds to the two entries of $1/2$ in the third matrix of \fig{ChessFig}.
In short, when the environment finds out about the system state, it decoheres.

\subsubsection{Effect of $\Hso$: decreasing entropy}

Whereas $\Hoe$ typically causes the apparent 
entropy of the object to increase,
$\Hso$ typically causes it to decrease.
\Fig{ChessFig} illustrates the case of an ideal measurement, where
the subject starts out in the state $\noobs$
and $\Hso$ is of such a form that 
the subject gets perfectly correlated with the object.
In the language of \eq{IdefEq},
an ideal measurement is a type of 
communication where the mutual information 
$I_{so}$ between the subject and object systems
is increased to its maximum possible value\cite{EverettBook}.
Suppose that the measurement is caused by $\Hso$ becoming large during 
a time interval so brief that we can neglect the effects of 
$\Hsubj$ and $\Hobj$. The joint subject$+$object density matrix
$\rso$ then evolves as $\rso\mapsto U\rso U^\dagger$,
where $U\equiv\exp\left[-i\int\Hso dt\right]$.
If observing $\up$ makes the subject happy and 
$\down$ makes the subject sad, then we have
$U\noup=\upup$ and $U\nodown=\downdown$.
The state given by \eq{DecoherenceExampleEq}
would therefore evolve into
\beqa{DecreasingEntropyEq}
\robj
&=&{1\over 2}U(\noobs\noobsbra)\tensormult(\up\upbra + \down\downbra)U^\dagger\\\nonumber
&=&{1\over 2}(U\noup\noupbra U^\dagger + U\nodown\nodownbra U^\dagger\\\nonumber
&=&{1\over 2}(\upup\upupbra + \downdown\downdownbra) = {1\over 2}\left(\rhohappy+\rhosad\right),
\eeqa
as illustrated in \fig{ChessFig}, where $\rhohappy\equiv\upup\upupbra$ and $\rhosad\equiv\downdown\downdownbra$.
This final state contains a mixture of 
two subjects, corresponding to definite but opposite 
knowledge of the object state.
According to both of them, the entropy of the object has
decreased from one bit to zero bits.
As mentioned above, there is a separate object density matrix $\robj$ corresponding to each of these two observers. 
Each of these two observers picks out her density matrix by conditioning the density matrix of \eq{DecreasingEntropyEq} on her subject degrees of freedom, \ie, the density matrix of the happy one 
is $\rhohappy$ and that of the other one is $\rhosad$. 
These are what Everett termed the ``relative states'' \cite{EverettBook}, except that we are expressing them in terms of density matrices rather than wavefunctions.
In other words, a subject by definition has zero entropy at all times, subjectively knowing her state perfectly.
Related discussion of the conditioning operation is given in \cite{brain,Nomura11}.

In many experimental situations, this projection step in defining the object density matrix corresponds to the familiar textbook process of quantum state preparation. 
For example, suppose an observer wants to perform a quantum measurement on a spin 1/2 silver atom in the state $\up$.
To obtain a silver atom prepared in this state, she can simply perform the measurement of one atom, introspect,  and 
if she finds that she is in state $\upobs$, then she know that her atom is prepared in the desired state $\up$ --- otherwise she discards it and tries again with other atom until she succeeds.
Now she is ready to perform her experiment.

In cosmology, this state preparation step is often so obvious that it is easy to overlook.
Consider for example the state illustrated in \fig{ExpandingUniverseSplitFig} and ask yourself what density matrix you should use to make predictions for your own future cosmological observations. 
All experiments you can ever perform are preceded by you introspecting and implicitly confirming that you are not in one of the stillborn galaxy-free wavefunction branches that failed to inflate. Since those dead branches are thoroughly decohered from the branch that you are in, they are completely irrelevant to predicting your future, and it would be a serious mistake not to discard their contribution to the density matrix of your universe.
This conditionalization is analogous to the use of conditional probabilities when making predictions in classical physics.
If you are playing cards, for example, the probabilistic model that you make for your opponents hidden cards reflects your knowledge of your own cards; you do not consider shuffling outcomes where you were dealt different cards than those you observe.

Just as decoherence can be partial, when $\langle\epsilon_j|\epsilon_i\rangle\ne 0$, so can measurement, so let us now derive how observation 
changes the density matrix also in the most general case.
Let $\ket{s_i}$ denote the basis states that the subject can perceive --- as discussed above, these must be robust to decoherence, and will for the case of a human observer correspond to ``pointer states'' \cite{ZurekHabibPaz93} of certain degrees of freedom of her brain.
Just as in the decoherence section above, let us consider general interactions that leave the object unchanged, \ie, such that the unitary dynamics is $\U\equiv e^{-i\Hso t}$ during the observation and
merely changes the 
subject state in a way that depends on the object state $\ket{o_i}$, say from an initial state $\ket{s_0}$ into some final state $\ket{\sigma_i}$:
\beq{ObsAssumptionEq}
U\ket{s_0}\ket{o_i}=\ket{\sigma_i}\ket{o_i}.
\eeq
This means that an initial density matrix $\rho=\ket{s_0}\bra{s_0}\tensormult\rho_o$ of the subject-object system, where
$\rho_o=\sum_{ij}\bra{o_i}\rho_o\ket{o_j}\ket{o_i}\bra{o_j}$,
will evolve as 
\beqa{ObsDerivationEq1}
\rho&\mapsto&U\rho\U^\dagger = U\ket{s_0}\bra{s_0}\rho_o\U^\dagger\nonumber\\
       &=&\sum_{ij}\bra{o_i}\rho_o\ket{o_j}U\ket{s_0}\ket{o_i}\bra{s_0}\bra{o_j}\U^\dagger\nonumber\\
       &=&\sum_{ij}\bra{o_i}\rho_o\ket{o_j}\ket{\sigma_i}\ket{o_i}\bra{\sigma_j}\bra{o_j}.
\eeqa
Since the subject will decohere rapidly, on a timescale much shorter than that on which subjective perceptions change, 
we can apply the decoherence formula~\eqn{vonNeumannReductionEq} to this expression with $P_i=\ket{s_i}\bra{s_i}$, which gives
\beqa{ObsDerivationEq2}
\rho&\mapsto&\sum_k P_k \rho P_k = \sum_k\ket{s_k}\bra{s_k}\rho\ket{s_k}\bra{s_k}\nonumber\\
       &=&\sum_{ijk}\bra{o_i}\rho_o\ket{o_j}\bra{s_k}\sigma_i\rangle\bra{\sigma_j}s_k\rangle \ket{s_k}\bra{s_k}\tensormult\ket{o_i}\bra{o_j}\nonumber\\
        &=&\sum_k  \ket{s_k}\bra{s_k}\tensormult\rho_o^{(k)},
\eeqa
where 
\beqa{RelativeStateEq}
\rho_o^{(k)}&\equiv&\sum_{ij}\bra{o_i}\rho_o\ket{o_j}\bra{s_k}\sigma_i\rangle\bra{\sigma_j}s_k\rangle \ket{o_i}\bra{o_j}\nonumber\\
	            &=& \sum_{ij}\P_i\rho_o\P_j\bra{s_k}\sigma_i\rangle\bra{s_k}\sigma_j\rangle^* 
\eeqa
is the (unnormalized) density matrix that the subject perceiving $\ket{s_k}$ will experience.
\Eq{ObsDerivationEq2} thus describes a sum of decohered components, each of which contains the subject in a
pure state $\ket{s_k}$. 
For the version of the subject perceiving $\ket{s_k}$, the correct object density matrix to use for all its future predictions is therefore
$\rho_o^{(k)}$ appropriately re-normalized to have unit trace:
\beqa{ObservationRhoEq}
\rho_o&\mapsto&{\rho_o^{(k)}\over\tr \rho_o^{(k)}} 
= { \sum_{ij}\P_i\rho_o\P_j\bra{s_k}\sigma_i\rangle\bra{s_k}\sigma_j\rangle^*\over  \sum_{i}\tr\rho_o\P_i|\bra{s_k}\sigma_i\rangle|^2  }\nonumber\\
&=&{\PP_k\rho\PP_k^\dagger\over\tr\PP_k\rho\PP_k^\dagger}, 
\eeqa
where
\beq{PiDefEq}
 \PP_k=\sum_i  \langle s_k|\sigma_i\rangle\P_i.
\eeq
This can be thought of as a generalization of Everett's so-called relative state from wave functions to density matrices and from complete to partial measurements. It can also be thought of as a generalization of Bayes' Theorem from the classical to the quantum case: just like the classical Bayes' theorem shows how to update an observer's classical description of something (a probability distribution) in response to new information, the quantum Bayes' theorem shows how to update an observer's quantum description of something (a density matrix).

We recognize the denominator $\tr\rho_o^{(k)}=\sum_{i}\bra{o_i}\rho_o\ket{o_i}|\bra{s_k}\sigma_i\rangle|^2$ as the standard expression for the probability that the subject will perceive $\ket{s_k}$.
Note that the same final result in \eq{ObservationRhoEq} can also be computed directly from \eq{ObsDerivationEq1} without invoking decoherence, 
as $\rho_o\mapsto \bra{s_k}\rho\ket{s_k}/\tr\bra{s_k}\rho\ket{s_k}$, so the role of decoherence lies merely in clarifying why this is the 
correct way to compute the new $\rho_o$.

To better understand \eq{ObservationRhoEq}, let us consider some simple examples:
\begin{enumerate}
\item If  $\langle s_i|\sigma_j\rangle=\delta_{ij}$, then we have a perfect measurement in the sense that the subject learns the exact object state, and 
\eq{ObservationRhoEq} reduces to $\rho_o\mapsto P_k$, \ie,. the observer perceiving $\ket{s_k}$ knows that the object is in its $k^{\rm th}$ eigenstate.
\item If  $\ket{\sigma_i}$ is independent of $i$, then no information whatsoever has been transmitted to the subject, and 
\eq{ObservationRhoEq} reduces to $\rho_o\mapsto\rho_o$, \ie, nothing changes.
\item If for some subject state $k$ we have $\langle s_i|\sigma_j\rangle=1$ for some group of $j$-values, vanishing otherwise, 
then the observer knows only that the object state is in this group (this can happen if $\Hso$ has degenerate eigenvalues).
\Eq{ObservationRhoEq} then reduces to
${\PP_k\rho\PP_k\over\tr \rho\PP_k}$, where $\PP_k$ is the projection operator onto this group of states.
\end{enumerate}

\subsubsection{Entropy and information}

In summary, we see that the object decreases its entropy
when it exchanges information with the subject and increases it
when it exchanges information with the environment.
Since the standard phrasing of the second law of thermodynamics is focused on the case where interactions with the observer are unimportant, we can rephrase it in a more nuanced way that explicitly acknowledges this caveat:

\smallskip\smallskip
\centerline{\framebox{\parbox{7cm}{
{\bf Second law of thermodynamics:\\} {\it The object's entropy can't decrease unless it interacts with the subject.}
}}}
\smallskip\smallskip

We can also formulate an analogous law that focuses on decoherence and ignores the observation process:

\smallskip\smallskip
\centerline{\framebox{\parbox{7cm}{
{\bf Another law of thermodynamics:\\} {\it The object's entropy can't increase unless it interacts with the environment.}
}}}
\smallskip\smallskip

In Appendix~\ref{EntropyProofSec}, we prove the first version and clarify the mathematical status and content of the second version.
Note that for the above version of the second law, we are restricting the interaction with the environment to be of a form
of \eq{DecoAssumptionEq}, \ie, to be such that it does not alter the state of the system, merely transfers information about it to the environment.
In contrast, if general object-environment interactions $\Hoe$ are allowed, then there are no restrictions on how the object entropy can change: for example, there is always an interaction that simply exchanges the state of the object with the state of part of the environment, and if the latter is pure, this interaction will decrease the object entropy to zero. More physically interesting examples of entropy-reducing object-environment interactions include dissipation (which can in some cases purify a high-energy mixed state to closely approach a pure ground state) and error correction (for example, where a living organism reduces its own entropy by absorbing particles with low entropy from the environment, performing unitary error correction that moves part of its own entropy to these particles, and finally dumping these particles back into the environment).

In regard to the other law of thermodynamics above, note that it is merely {\it on average} that interactions with the object cannot increase the entropy
(because of Shannon's famous result that the entropy gives the average number of bits required to specify an outcome). For certain individual measurement outcomes, observation can sometimes increase entropy --- we will see an explicit example of this in the next section.

For a less cosmological example, consider helium gas in a thermally insulated box, starting off with the gas particles in a zero-entropy coherent state, where each atom is in a rather well-defined position. There are positive Lyapunov exponents in this system because the momentum transfer in atomic collisions is sensitive to the impact parameter, so before long, chaotic dynamics has placed
every gas particle in a superposition of being everywhere in a box --- indeed, in a superposition of being all over phase space, with a Maxwell-Boltzmann distribution. 
If we define the object to be some small subset of the helium atoms and call the rest of the atoms the environment, then the object entropy $S_o$ will be high (corresponding to a roughly thermal density matrix $\rho_o\propto e^{-H/kT}$) even though the the total entropy $S_{oe}$ remains zero; the difference between these two entropies reflects the information that the environment has about the object via quantum entanglement as per \eq{IdefEq}.
In classical thermodynamics, the only way to reduce the entropy of a gas is to invoke Maxwell's demon.
Our formalism provides a different way to understand this: the entropy decreases if you yourself are the demon, obtaining information about the individual atoms that constitute the object.

\section{Application to the inflationary entropy problem}
\label{EntropyProbSec}

\begin{figure}[pbt]
\centerline{\includegraphics[width=87mm]{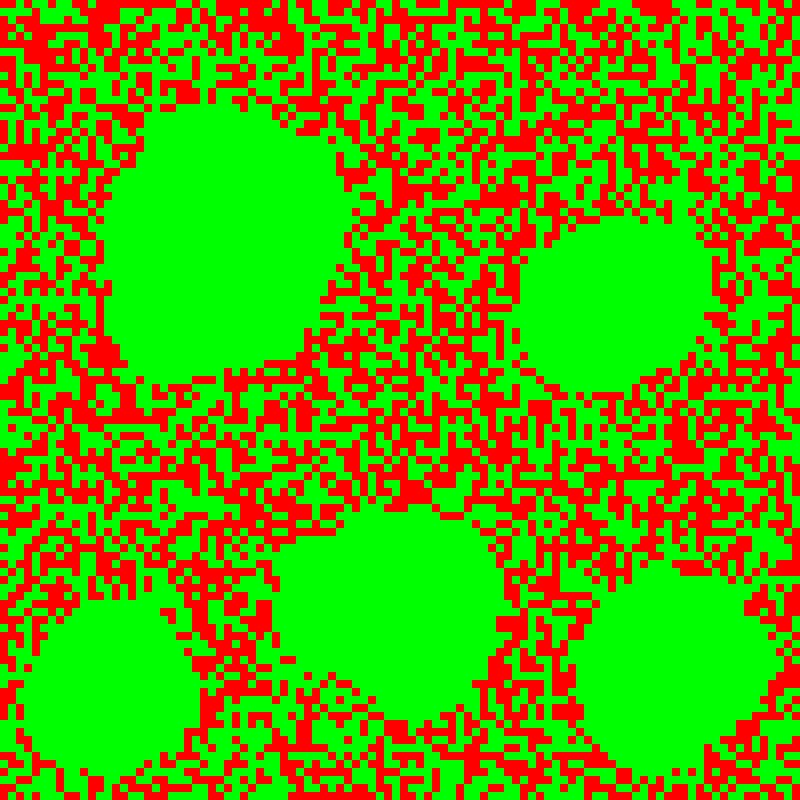}}
\bigskip
\caption{Our toy model involves a pixelized space where pixels are habitable (green/light grey) or uninhabitable (red/dark grey) at random with probability 50\%, except inside large contiguous inflationary patches where all pixels are habitable.  
}
\label{PatchesFig}
\end{figure}

\subsection{A classical toy model}
\label{ClassicalSolutionSec}

To build intuition for the effect of observation on entropy in inflationary cosmology, let us consider the simple toy model illustrated in \fig{PatchesFig}.
This model is purely classical, but we will show below how the basic conclusions generalize to the quantum case as well.
We will also see that the qualitative conclusions remain valid when this unphysical toy model is replaced by realistic inflation scenarios.

Let us imagine an infinite space pixelized into discrete voxels of finite size, each of which can be in only two states.
We will refer to these two states as {\it habitable} and  {\it uninhabitable}, and in \fig{PatchesFig}, they are colored 
green/light grey and red/dark grey, respectively.
We assume that some inflation-like process has created large habitable patches in this space, which fill a fraction $f$ of the total volume, and that the rest of space has a completely random state where each voxel is habitable with  50\% probability, independently of the other voxels.

Now consider a randomly selected region (which we will refer to as a ``universe'' by analogy with our Hubble volume) of this space, lying either completely inside an inflationary patch or completely outside the inflationary patches --- almost all regions much smaller than the typical inflationary patch will have this property.
Let us number the voxels in our region in some order $1,2,3,...$, and let us represent each state by a string of zeroes and ones denoting habitable and uninhabitable, where a 0 in the $i^{\rm th}$ position means that the $i^{\rm th}$ voxel is habitable.
For example, if our region contains 30 voxels, then ``000000000000000000000000000000" denotes the state where the whole region is habitable, whereas  ``101101010001111010001100101001'' represents a rather typical non-inflationary state. Finally, we label each state by an integer $i$ which is simply its bit string interpreted as a binary number.

 Letting $n$ denote the number of voxels in our region, there are clearly $2^n$ possible states $i=0,...,2^n-1$ that it can be in.
By our assumptions, the probability $p_i$ that our region is in the $i^{\rm th}$ state (denoted $A_i$) is 
\beq{ProbDistEq}
p_i\equiv P(A_i)=\left\{
\begin{tabular}{lll}
$f+(1-f)2^{-n}$	&if&$i=0$,\\
$(1-f)2^{-n}$	&if&$i>0$,
\end{tabular}
\right.
\eeq
\ie, 
there is a probability $f$ of being in the $i=0$ state because inflation happened in our region, plus a small probability
$2^{-n}$ of being in any state in case inflation did not happen here.

Now suppose that we decide to measure $b$ bits of information by observing the state of the first $b$ voxels.
The probability $P(H)$ that they are all habitable is simply the total probability of the first $2^{n-b}$ states, \ie,
\beqa{PhabitableEq}
P(H)&=&\sum_{i=0}^{2^{n-b}-1} p_i= f+(1-f)2^{-n} + (2^{n-b}-1)(1-f)2^{-n}\nonumber\\
                    &=&f+(1-f)2^{-b},
\eeqa
independent of the number of voxels $n$ in our region. This result is easy to interpret: 
either we are in an inflationary region (with probability $f$), in which case these $b$ voxels are all habitable, 
or we are not (with probability $1-f$), in which case they are all habitable with probability $2^{-b}$.

If we find that these $b$ voxels are indeed all habitable, then 
using the standard formula for conditional probabilities, we obtain the following 
revised probability distribution for the
state of our region:
\beqa{PconditionalEq}
p_i^{(b)}&\equiv&P(A_i|H)={P(A_i \& H)\over P(H)}\nonumber\\
               &=&\left\{
\begin{tabular}{lll}
${f+(1-f)2^{-n}\over f+(1-f)2^{-b}}$	&if&$i=0$,\\
${(1-f)2^{-n}\over f+(1-f)2^{-b}}$	&if&$i=1,...,2^{n-b}-1$,\\
$0$							&if&$i=2^{n-b},...,2^n-1$.
\end{tabular}
\right.
\eeqa
We are now ready to compute the entropy $S$ of our region given various degrees of knowledge about it, which is defined by the standard Shannon formula
\beq{ShannonEq}
S^{(b)}\equiv\sum_{i=0}^{2^n-1}h\left[p^{(b)}_i\right],\quad h(p)\equiv -p\log p,
\eeq
where, as mentioned, we use logarithms of base two so that the entropy has units of bits.
Consider first the simple case of no inflation, $f=0$. Then all non-vanishing probabilities reduce to $p_i^{(b)}=2^{b-n}$ and the entropy is simply
\beq{NoInflationEntropyEq}
S^{(b)}=n-b.
\eeq
In other words, the state initially requires $n$ bits to describe, one per voxel, and whenever we observe one more voxel, the entropy drops by one bit: the one bit of information we gain tells us merely about the state of the observed voxel, and tells us nothing about the rest of space since the other voxels are statistically independent.

More generally, substituting \eq{PconditionalEq} into \eq{ShannonEq} gives
\beq{FullEntropyEq}
S^{(b)} = h\left[{f+(1-f)2^{-n}\over f+(1-f)2^{-b}}\right] + \left(2^{n-b}-1\right)h\left[{(1-f)2^{-n}\over f+(1-f)2^{-b}}\right]. 
\eeq
As long as the number of voxels is large ($n\gg b$) and the 
inflated fraction $f$ is non-negligible ($f\gg 2^{-n}$), this entropy is accurately approximated by
\beqa{ApproxEntropyEq}
&&S^{(b)}\approx h\left[{f\over f+(1-f)2^{-b}}\right] + 2^{n-b} h\left[{(1-f)2^{-n}\over f+(1-f)2^{-b}}\right]\\
&&={n\over {2^b f\over 1-f} + 1} + {h(f)+2^{-b}h(1-f)\over  f+(1-f)2^{-b}} +\log\left[f+(1-f)2^{-b}\right] \nonumber.
\eeqa
The sum of the last two terms is merely an $n$-independent constant of order unity which approaches zero
as we observe more voxels (as $b$ increases), so in this limit, 
\eq{ApproxEntropyEq} reduces to simply 
\beq{ApproxEntropyEq2}
S^{(b)}\approx{(f^{-1}-1) n\over 2^b}.
\eeq
For the special case $f=1/2$ where half the volume is inflationary, \eq{ApproxEntropyEq} reduces to the more accurate result
\beq{ApproxEntropyEq3}
S^{(b)}\approx{n\over 2^b+1} + \log[1+2^{-b}]
\eeq
without approximations.

\begin{figure}[pbt]
\centerline{\includegraphics[width=87mm]{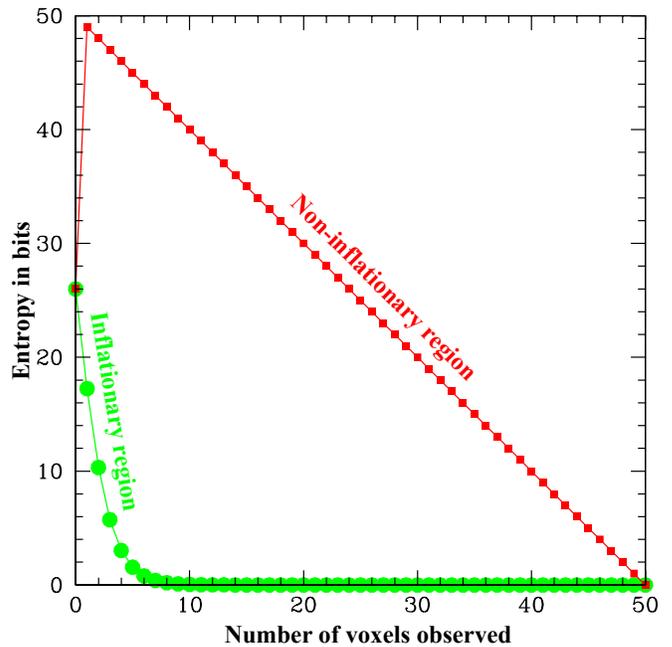}}
\bigskip
\caption{How observations change the entropy for an inflationary fraction $f=0.5$.
If successive voxels are all observed to be habitable, the entropy drops roughly exponentially in accordance with 
\eq{ApproxEntropyEq3} (green/grey dots). 
If the first voxel is observed to be uninhabitable, thus establishing that we are in a non-inflationary region, then
the entropy instead shoots up to the line of slope $-1$ given by \eq{NoInflationEntropyEq} (grey/red squares). More generally, we observe $b$ habitable voxels and then one uninhabitable one, the entropy first follows the dots, then jumps up to the squares, then follows the squares downward regardless of what is observed thereafter.
This figure illustrates the case with $n=50$ voxels --- although 
$n\sim 10^{120}$ is more relevant to our actual universe, the drop toward zero of the green curve would be too fast to be visible in the such a plot.
}
\label{EntropyFig}
\end{figure}

Comparing \eq{NoInflationEntropyEq} with either of the last two equations, we notice quite a remarkable difference, which is illustrated in \fig{EntropyFig}: 
in the inflationary case, the entropy decreases not linearly (by one bit for every bit observed), but exponentially!
This means that in our toy inflationary universe model, if an observer looks around and finds that even a tiny nearby volume is habitable, this dramatically reduces the entropy of her universe. For example, if $f=0.5$ and there are $10^{120}$ voxels, then the initial entropy is about $10^{120}$ bits, and observing merely 400 voxels (less than a fraction $10^{-117}$ of the volume) to be habitable brings this huge entropy down to less than one bit.

How can observing a single voxel have such a large effect on the entropy?
The answer clearly involves the long-range correlations induced by inflation, whereby this single voxel carries information about whether inflation occurred or not in all the other voxels in our universe. 
If we observe $b\gg-\log f$ habitable voxels, it is exponentially unlikely that we are not in an inflationary region. We therefore know with virtual certainty that the voxels that we will observe in the future are also habitable. Since our uncertainty about the state of these voxels has largely gone away, the entropy must have decreased dramatically, as \eq{ApproxEntropyEq2} confirms.

To gain more intuition for how this works, consider what happens if we instead observe the first $b$ voxels to be {\it uninhabitable}.
Then \eq{PconditionalEq} instead makes all non-vanishing probabilities $p_i=2^{b-n}$, and we recover 
\eq{NoInflationEntropyEq} even when $f\ne 0$. 
Thus observing merely the first voxel to be uninhabitable causes the entropy to dramatically {\it increase},
from $(1-f)n$ to $n-1$, roughly doubling if $f=0.5$.
We can understand all this by recalling Shannon's famous result that the entropy gives the average number of bits required to specify an outcome.
 If we know that our universe is not inflationary, then we need a full $n$ bits of information to specify the state of the $n$ voxels, since they are all independent. 
If we know that our universe {\it is} inflationary, on the other hand, then we know that all voxels are habitable, and we need no further information. Since a a fraction $(1-f)$ of the universes are non-inflationary, we thus need $(1-f)n$ bits on  average.
Finally, to specify whether it is inflationary or not, we need 1 bit of information if $f=1/2$ and more generally the slightly smaller amount $h(f)+h(1-f)$, which is the entropy of a two-outcome distribution with probabilities $f$ and $1-f$. 
The average number of bits needed to specify a universe is therefore
\beq{BitsNeededEq}
S^{(0)} \approx (1-f)n + h(f)+h(1-f),
\eeq 
which indeed agrees with \eq{ApproxEntropyEq} when setting $b=0$.

In other words, the entropy of our universe before we have made any observations is the average of a very large number and a very small number, corresponding to inflationary and non-inflationary regions. As soon as we start observing, this entropy starts leaping towards one of these two numbers, reflecting our increased knowledge of which of the two types of region we inhabit.

Finally, we note that the success in this inflationary explanation of low entropy does not require an extreme anthropic selection effect where life is {\it a priori} highly unlikely; contrariwise, the probability that our entire universe is habitable is simply $f$, and the effect works fine also when $f$ is of order unity.

\subsection{The quantum case}

To build further intuition for the effect of observation on entropy, let us generalize our toy model to include quantum mechanics.
We thus upgrade each voxel to a 2-state quantum system, with two orthogonal basis states denoted
$\ket{0}$ (``habitable'') and $\ket{1}$ (``uninhabitable'').
The Hilbert space describing the quantum state of an $n$-voxel region thus has $2^n$ dimensions.
We label our $2^n$ basis states by the same bit strings as earlier, so 
the state of the 30-voxel example given in \Sec{ClassicalSolutionSec} above would be written
\beq{SampleKetEq}
 \ket{\psi_i} = \ket{101101010001111010001100101001},
\eeq
corresponding to basis state $i= 759\>669\>545$.
If the region is inflationary, all its voxels are habitable, so its density matrix is 
\beq{rhoYesEq}
\rho_{\rm yes}=\ket{000\dots 0}\bra{000\dots0}
\eeq
If it is not inflationary, then we take each voxel to be in the mixed state 
\beq{rhostarDefEq}
\rho_*= \frac{1}{2}\left[\ket{0}\bra{0}+\ket{1}\bra{1}\right],
\eeq
independently of all the other voxels, and the density matrix $\rho_{\rm no}$ of the whole region is simply a tensor product of $n$ such 
single-voxel density matrices.
In the general case that we wish to consider, there is a probability $f$ that the region is inflationary, so the full density matrix is
\beqa{RegionInitialStateEq}
\rho&=&f \rho_{\rm yes} + (1-f)\rho_{\rm no}\\
       &=&f \ket{000\dots 0}\bra{000\dots0} + (1-f)\rho_*\tensormult\rho_*\tensormult\rho_*\tensormult\dots\rho_* \nonumber
\eeqa
Expanding the tensor products, it is easy to show that we get $2^n$ different terms, and that this full density matrix can be rewritten in the form 
\beq{rhoEq2}
\rho = \sum_{i=0}^{2^n-1} p_i\ket{\psi_i}\bra{\psi_i},
\eeq
where $p_i$ are the probabilities given by \eq{ProbDistEq}.

Now suppose that we, just as in the previous section, decide to measure $b$ bits of information by observing the state of the first $b$ voxels and find them all to be habitable.
To compute the resulting density matrix $\rho^{(b)}$, we thus condition on our observational results using \eq{ObservationRhoEq} with the projection matrix $\P=\ket{0...0}\bra{0...0}$, with $b$ occurrences of $0$ inside each of the two brackets, obtaining
\beq{QuantumConditionalEq}
\rho^{(b)} = {\P\rho\P\over\tr\P\rho\P}.
\eeq
Substituting \eq{rhoEq2} into this expression and performing some straightforward algebra gives
\beq{rhoConditionalEq}
\rho^{(b)} = \sum_{i=0}^{2^n-1} p_i^{(b)}\ket{\psi_i}\bra{\psi_i},
\eeq
where $p_i^{(b)}$ are the probabilities given by \eq{PconditionalEq}.
We can now compute the quantum entropy $S$ of our region, which is defined by the standard von Neuman formula
\beq{vonNeumannEq}
S^{(b)}\equiv \tr h\left[\rho^{(b)}\right],\quad h(\rho)\equiv -\rho\log\rho,
\eeq
where we again use logarithms of base two so that the entropy has units of bits.
This trace is conveniently evaluated in the $\ket{\psi_i}$-basis where \eq{rhoConditionalEq} shows that
the density matrix $\rho^{(b)}$ is diagonal, reducing the entropy to the sum
\beq{ShannonEq2}
S^{(b)}\equiv\sum_{i=0}^{2^n-1}h\left[p^{(b)}_i\right].
\eeq
Comparing this with \eq{ShannonEq}, we see that this result is identical to the one we derived for the classical case.
In other words, all conclusions we drew in the previous section generalize to the quantum-mechanical case as well.

\subsection{Real-world issues}

Although we repeatedly used words like ``inflation'' and ``inflationary'' above, our toy models of course contained no inflationary physics whatsoever.
For example, real eternal inflation tends to produce a messy spacetime with significant curvature on scales far beyond the cosmic particle horizon, not simply large uniform patches embedded in Euclidean space\footnote{It is challenging to quantify the inflationary volume fraction $f$ in such a messy spacetime, but as we saw above, this does not affect the qualitative conclusions as long as $f$ is not exponentially small --- which appears unlikely given the tendency of eternal inflation to dominate the total volume produced.}, 
and real inflation has quantum field degrees of freedom that are continuous rather than simple qubits. However, it is also clear that our central result regarding exponential entropy reduction has a very simple origin that is independent of such physical details:  {\it long-range entanglement}. In other words, the key was simply that the state of a small region could sometimes reveal the state of a much larger region around it
(in our case, local smoothness implied large-scale smoothness). 
This allowed a handful of measurements in that small region to, with a non-negligible probability, provide a massive entropy reduction by revealing that the larger region was in a very simple state.
We saw that the result was so robust that it did not even matter whether this long-range entanglement was classical or quantum-mechanical.

It is not merely inflation that produces such long-range entanglement, but {\it any} process that spreads rapidly outward from scattered starting points.
To illustrate this robustness to physics details, consider the alternative example where \fig{PatchesFig} is a picture of bacterial colonies growing in a Petri dish:
the contiguous spread of colonies creates long-range entanglement, so that observing a small patch to be colonized makes it overwhelmingly likely that a much larger region around it is colonized.  
Similarly, if you discover that a drop of milk tastes sour, it is extremely likely that a much larger volume (your entire milk carton) is sour.
A random bacterium in a milk carton should thus expect the entire carton to be sour just like a random cosmologists in a habitable post-inflationary patch of space should expect her entire Hubble volume to be post-inflationary.

\section{Discussion}
\label{DiscussionSec}

In the context of unitary cosmology, we have investigated the time-evolution of the density matrix with which an observer describes a quantum system, focusing on the processes of decoherence and observation and how they change entropy.
Let us now discuss some implications of our results for inflation and quantum gravity research.

\subsection{Implications for inflation}

Although inflation has emerged as the most popular theory for what happened early on, bolstered by improved measurements involving the cosmic microwave background and other cosmological probes, the case for inflation is certainly not closed.
Aside from issues  to do with measurements \cite{Copi10} and competing theories 
\cite{PreBigBang93, Ekpyrotic01,BrandenbergerStringGas09}, there are at least four potentially serious problems with its theoretical foundations, which are arguably interrelated:
\begin{enumerate}
\item{The entropy problem}
\item{The measure problem}
\item{The start problem}
\item{The degree-of-freedom problem}
\end{enumerate}
Since we described the entropy problem in the introduction, let us now briefly discuss the other three.
Please note that we will not aim or claim to solve any of these three additional problems in the present paper, merely to highlight them and describe additional difficulties related to the degree-of-freedom problem. 

\subsubsection{The measure problem}

Inflation is generically eternal, producing a messy spacetime with infinitely many post-inflationary pockets separated by regions that inflate forever  \cite{Vilenkin83,Starobinsky84,LindeBook}.
These pockets together contain an infinite volume and infinitely many particles, stars and planets.
Moreover, certain observable quantities like the density fluctuation amplitude that we have observed to be $Q\sim 2\times 10^{-5}$ in our part of spacetime \cite{Q,wmap7} take different values in different places.\footnote{$Q$ depends on how the inflaton field rolled down its potential, so for a 1-dimensional potential with a single minimum, $Q$ is generically different in regions where the field rolled from the left and from the right. If there potential has more than one dimension, there is a continuum of options, and if there are multiple minima, 
there is even the possibility that other effective parameters (physical ``constants'') may differ between different minima, as in the string theory landscape scenario 
 \cite{Bousso00,Feng00,KKLT03,Susskind03,AshikDouglas04}.}
Taken together, these two facts create what has become known as the inflationary ``measure problem'' 
\cite{Linde95,multiverse4wheeler,conditionalization,inflation,Freivogel05,Garriga05,Easther05,axions,Aguirre06a, Aguirre06b,Bousso06,GibbonsTurok08,HartleHertog07,Page08,deSimone08,Linde09,Bousso10,Linde10, Page10,Vilenkin11,Salem11,GuthVanchurin11}:
the predictions of inflation for certain observable quantities are not definite numbers, merely probability distributions, and we do not yet know  how to compute these distributions.

The failure to predict more than probability distributions is of course not a problem {\it per se}, as long as we know how to compute them (as in quantum mechanics). 
In inflation, however, there is still no consensus around any unique and well-motivated framework for computing such probability distributions despite a major community effort in recent years. The crux of the problem is that when we have a messy spacetime with infinitely many observers who subjectively feel like you, any procedure to compute the fraction of them who will measure say one $Q$-value rather than another will depend on the order in which you count them, just as the fraction of the integers that are even depends on the order in which you count them \cite{inflation}. 
There are infinitely many such observer ordering choices, many of which appear reasonable yet give 
manifestly incorrect predictions 
\cite{inflation,Aguirre06a,Page08,Vilenkin11,GuthVanchurin11}, 
and despite promising developments, the measure problem remains open.
A popular approach is to count only the finite number of observers existing before a certain time $t$ and then letting $t\to\infty$, but this procedure has turned out to be extremely sensitive to the choice of time variable $t$ in the spacetime manifold, with no obviously correct choice 
\cite{inflation,Aguirre06a,Page08,Vilenkin11,GuthVanchurin11},
The measure problem has eclipsed and subsumed the so-called fine tuning problem, in the sense that even the rather special inflaton potential shapes that are required to match observation can be found in many parts of the a messy multidimensional inflationary potential suggested by the string landscape scenario with its $10^{500}$ or more 
distinct minima \cite{Bousso00,Feng00,KKLT03,Susskind03,AshikDouglas04}, so the question shifts from asking why our inflaton potential is the way it is to asking what the probability is of finding yourself in different parts of the landscape.

In summary, until the measure problem is solved, inflation strictly speaking cannot make any testable predictions at all, thus failing to qualify as a scientific theory in Popper's sense.


\subsubsection{The start problem}

Whereas the measure problem stems from the end of inflation (or rather the lack thereof), a second problem stems from the beginning of inflation. 
As shown by Borde, Guth \& Vilenkin \cite{BordeGuthVilenkin03}, 
inflation must have had a beginning,  \ie, cannot be eternal to the past (except for the loophole described in \cite{AguirreGratton02, AguirreGratton03}),
so inflation fails to provide a complete theory of our origins, and needs to be supplemented with a theory of what came before.
(The same applies to various ekpyrotic and cyclic universe scenarios \cite{BordeGuthVilenkin03}.)

The question of what preceded inflation is wide open, with proposed answers including
quantum tunneling from nothing \cite{Vilenkin83,HartleHawking83}, 
quantum tunneling from a ``pre-big-bang'' string perturbative vacuum \cite{PreBigBang92,PreBigBang01}
and quantum tunneling from some other non-inflationary state. 
Whereas some authors have argued that eternal inflation makes predictions that are essentially independent of how inflation started, others have argued that this is not the case \cite{Garriga06,Aguirre09,Freivogel09}. Moreover, there is no quantitative agreement between the probabilities predicted by different scenarios, some of which even differ over the sign of a huge exponent.

The lack of consensus about the start of inflation not only undermines claims that inflation provides a final answer, but also calls into question whether some of the claimed successes of inflation really are successes. 
In the context of the above-mentioned entropy problem, some have argued that tunneling into the state needed to start inflation is just as unlikely as tunneling straight into the current state of our universe \cite{CarrollBook,CarrollTam10}, whereas others have argued that inflation still helps by 
reducing the amount of mass that the quantum tunneling event needs to generate \cite{AlbrechtSorbo94}.

\subsubsection{The degree-of-freedom problem}

A third problem facing inflation is to quantum-mechanically understand what happens when a region of space is expanded indefinitely. 
We discuss this issue in detail in Appendix~\ref{SnapSec} below, and provide merely a brief summary here.
Quantum gravity considerations suggest that the number of quantum degrees of freedom $N$ in a comoving volume $V$ is finite.
If $N$ increases as this volume expands, then we need an additional law of physics that specifies when and where new degrees of freedom are created, and into what quantum states they are born. If $N$ does not increase, on the other hand, life as we know it may eventually be destroyed in a ``Big Snap'' when the increasingly granular nature of space begins to alter our effective laws of particle physics, much like a rubber band cannot be stretched indefinitely before the granular nature of its atoms cause our continuum description of it to break down. 
Moreover, in the simplest scenarios where the number of observers is proportional to post-inflationary volume, such Big Snap scenarios are already ruled out by  dispersion measurements using gamma ray bursts.
In summary, none of the three logical possibilities for the number of quantum degrees of freedom $N$ (it is infinite, it changes, it stays constant) is problem free.

\subsubsection{The case for inflation: the bottom line}

In summary, the case for inflation will continue to lack a rigorous foundation until the measure problem, the start problem and the degree-of-freedom problem have been solved, so until then, we cannot say for sure whether inflation solves the entropy problem and adequately explains our low observed entropy.
However, our results have shown that inflation certainly makes things better. 
We have seen that claims to the contrary are based on an unjustified neglect of the density matrix  conditioning requirement (the third dynamical equation in Table~1), thus conflating the entropy of the full quantum state with the entropy of subsystems.

Specifically, we have showed that by producing a quantum state with long-range entanglement, inflation creates a situation where observations can cause an exponential decrease in entropy, so that merely a handful of quantum measurements can bring the entropy for our observable universe down into the low range that we in fact observe.
This means that if we assume that sentient observers require at least a small volume (say enough to fit a few atoms) of low temperature ($\ll 10^{16}$\,GeV), then
almost all sentient observers will find themselves in a post-inflationary low-entropy universe, and we humans have no reason to be surprised that we do so as well.

\subsection{Implications for quantum gravity}

We saw above that  unjustified neglect of the density matrix conditioning requirement (the third dynamical equation in Table~1) can lead to incorrect conclusions about inflation. The bottom line is that we must not conflate the total density matrix with the density matrix relevant to us.
Interestingly, as we will now describe, this exact same conflation has led to various incorrect claims in the the literature about quantum gravity and dark energy, for example that dark energy is simply back-reaction from super-horizon quantum fluctuations.

\subsubsection{Is $G_{\mu\nu}\approx 8\pi G \expec{T_{\mu\nu}}$?}

Since we lack a complete theory of quantum gravity, we need some approximation in the interim for how quantum systems gravitate, generalizing 
the Einstein equation $G_{\mu\nu}= 8\pi G T_{\mu\nu}$ of General Relativity.
A common assumption in the literature is that to a good approximation,
\beq{TavgEq}
G_{\mu\nu}= 8\pi G \expec{T_{\mu\nu}},
\eeq
where $G_{\mu\nu}$ on the left-hand-side is the usual classical Einstein tensor specifying spacetime curvature, while 
$\expec{T_{\mu\nu}}$ on the right-hand-side denotes the expectation value of the quantum field theory operator $T_{\mu\nu}$,
\ie, $\expec{T_{\mu\nu}}\equiv\tr[\rho T_{\mu\nu}]$, where $\rho$ is the density matrix.
Indeed, this assumption is often (as in some of the examples cited below) made without explicitly stating it, as if its validity were self-evident.

So is the approximation of \eq{TavgEq} valid? It clearly works well in many cases, which is why it continues to be used.
Yet it is equally obvious that it cannot be universally valid.
Consider the the simple example of inflation with a quadratic potential starting out in a homogeneous and isotropic quantum state.
This state will qualitatively evolve as in \fig{ExpandingUniverseSplitFig}, into a quantum superposition of
many macroscopically different states, some of which correspond to a large semiclassical post-inflationary universe like ours (each with its planets {\protect\etc} in different places).
Yet since both the initial quantum state and the evolution equations have translational and rotational invariance, the final quantum state will too, which means that 
$\expec{T_{\mu\nu}}$ is homogeneous and isotropic.
But \eq{TavgEq} then implies that $G_{\mu\nu}$ is homogeneous and isotropic as well, \ie, that spacetime is exactly described by the Friedmann-Robertson-Walker metric. The easiest way to experimentally rule this out is to stand on your bathroom scale and note the gravitational force pulling you down.
In this particular branch of the wavefunction there is a planet beneath you, pulling you downward,
and it is irrelevant that there are other decohered branches of the wavefunction where the planet is instead above you, to your left, to your right, {\etc}, giving an average force of zero.  $\expec{T_{\mu\nu}}$ is position-independent for the quantum field density matrix corresponding to the total state, whereas the 
relevant density matrix is the one that is conditioned on your perceptions thus far, which include the observation that there is a planet beneath you.

The interesting question regarding \eq{TavgEq} thus becomes more nuanced: when exactly is it a good approximation?
In this spirit, \cite{Tsamis10} poses two questions:  {\it ``How unreliable are expectation values?''} and
{\it How much spatial variation should one expect?}
 We have seen above that the first step toward a correct treatment is to compute the density matrix conditioned on our observations
 (the third dynamic process in Table~1) and use this density matrix $\rho$ to describe the quantum state.
 Having done this, the question of whether \eq{TavgEq} is accurate basically boils down to the question of whether the quantum state is roughly ergodic, 
 \ie, whether a small-scale {\it spatial} average of a typical classical realization is well-approximated by the quantum {\it ensemble} average 
 $\expec{T_{\mu\nu}}\equiv\tr[\rho T_{\mu\nu}]$.
This ergodicity tends to hold for many important cases, including the inflationary case where the quantum wavefunctional for the primordial fields in our Hubble volume is roughly Gaussian, homogeneous and isotropic \cite{multiverse4wheeler}.
Spatial averaging on small scales is relevant because it tends to have little effect on the gravitational field on larger spatial scales, which depends mainly on the large-scale mass distribution, not on the fine details of where the mass is located.
For a detailed modern treatment of small-scale averaging and its interpretation as ``integrating out'' UV degrees of freedom, see \cite{Baumann10}.
Since very large scales tend to be observable and very small scales tend to be unobservable, 
 a useful rule-of-thumb in many situations is ``condition on large scales,  trace out small scales".

In summary, the popular approximation of \eq{TavgEq} is accurate if both of these conditions hold:
\begin{enumerate}
\item The spatially fluctuating stress-energy tensor for a generic branch of the wavefunction can be approximated by its spatial average.
\item The quantum ensemble average can be approximated by a spatial average for a generic branch of the wavefunction.
\end{enumerate}

\subsubsection{Dark energy from superhorizon quantum fluctuations?}

The discovery that our cosmic expansion is accelerating has triggered a flurry of proposed theoretical explanations, most of which involve some form of substance or vacuum density dubbed {\it dark energy}. 
An alternative proposal that has garnered significant interest is that there is no dark energy, and that the accelerated expansion is instead due to gravitational back-reaction from inflationary density perturbations on scales much larger than our cosmic particle horizon \cite{Barausse05,Kolb05}.
This was rapidly refuted by a number of groups \cite{Flanagan05,HirataSeljak05, Geshnizjani05,IshibashiWald05}, 
and a related claim that superhorizon perturbations can explain away dark energy \cite{MartineauBrandenberger05} was rebutted by \cite{KumarFlanagan08}.
 
Although these papers mention quantum mechanics perfunctorily at best (which is unfortunate given that the origin of inflationary perturbations is a purely quantum-mechanical phenomenon),  a core issue in these refuted models is precisely the one we have emphasized in this paper: the importance of using the correct density matrix, conditioned on our observations, rather than a total density matrix that implicitly involves incorrect averaging --- either quantum ``ensemble'' averaging as in \eq{TavgEq} or spatial averaging.
For example, as explained in \cite{KumarFlanagan08}, a problem with the effective stress-energy tensor $\expec{T_{\mu\nu}}$ of \cite{MartineauBrandenberger05} is that it involves averaging over regions of space beyond our cosmological particle horizon, even though our observations are limited to our backward lightcone.

Such unjustified spatial averaging is the classical physics equivalent of unjustified use of the full density matrix in quantum mechanics: 
in both cases, we get correct statistical predictions only if we predict the future given what we know about the present.
Classically, this corresponds to using conditional probabilities, and quantum mechanically this corresponds to conditioning the density matrix using the bottom equation of Table~1 --- neither is optional. 
In classical physics, you shouldn't expect to feel comfortable in boiling water full of ice chunks just because the spatially averaged temperature is lukewarm.
In quantum mechanics, you shouldn't expect to feel good when entering water that's in a superposition of very hot and very cold.
Similarly, if there is no dark energy and the total quantum state $\rho$ of our spacetime corresponds to a superposition of states with different amplitudes for superhorizon modes,
then we shouldn't expect to perceive a single semiclassical spacetime that accelerates (as claimed for some models \cite{Barausse05,Kolb05}), but 
rather to perceive one of many semiclassical spacetimes from a decohered superposition, all of which decelerate.

Dark energy researchers have also devoted significant interest to so-called phantom dark energy, which has an equation of state $w<-1$ and can lead to a ``big rip'' a finite time from now, when the dark energy density and the cosmic expansion rate becomes infinite, ripping apart everything we know.
The same logical flaw that we highlighted above would apply to all attempts to derive such results by exploiting 
infrared logarithms in the equations for density and pressure \cite{OnemliWoodard2004} if they give $w<-1$ on scales much larger than our cosmic horizon, or more generally to talking about ``the equation of state of a superhorizon mode'' without carefully spelling out and justifying any averaging assumptions made.

 %


\subsection{Unitary thermodynamics and the Copenhagen Approximation}

In summary, we have analyzed cosmology assuming unitary quantum mechanics, using a tripartite partition into system, observer and environment degrees of freedom.
We have seen that this generalizes the second law of thermodynamics to {\it ``The system's entropy can't decrease unless it interacts with the observer, and it can't increase unless it interacts with the environment''}. Quantitatively, the system (``object'') density matrix evolves according to one of the three equations listed in Table~1 depending on whether the main interaction of the system is with itself, with the environment or with the observer. 
The key results in this paper follow from the third equation of Table~1, which gives the evolution of the quantum state under an arbitrary measurement or state preparation, and can be thought of as a generalization of the POVM (Positive Operator Valued Measure) formalism \cite{SudarshanMathewsRau60,Krauss83}.

Informally speaking, the entropy of an object decreases while you look at it and increases while you don't \cite{brain}. 
The common claim that entropy cannot decrease simply corresponds to the approximation of ignoring the subject in \fig{TrinityFig}, \ie, ignoring measurement.
Decoherence is simply a measurement that you don't know the outcome of, and measurement is simply entanglement, a transfer of quantum information about the system: 
the decoherence effect on the object density matrix (and hence the entropy)
is identical regardless of whether this measurement is performed by another person, a mouse, a computer or a single particle that encodes information about the system by bouncing off of it.\footnote{As described in detail, \eg, 
\cite{Zeh70,JZ85,ZehBook,Zurek09,SchlosshauerBook}, decoherence is not simply the suppression of 
off-diagonal density matrix elements in general, but rather the occurrence of this in the particular basis of relevance to the observer. This basis is in turn determined dynamically by decoherence of both the object \cite{Zeh70,JZ85,ZehBook,Zurek09,SchlosshauerBook} and the subject  \cite{brain,ZehBrain}.}
In other words, observation and decoherence both share the same first step, with another system obtaining information about the object --- the only difference is whether that system is the subject or the environment, \ie, whether the last step is conditioning or partial tracing:
\begin{itemize}
\itemsep0cm
\item {\it observation\hskip1.9mm = entanglement + conditioning}
\item {\it decoherence = entanglement + partial tracing}
\end{itemize}

Our formalism assumes only that quantum-mechanics is unitary and applies even to observers --- \ie, we assume that 
observers are physical systems too, whose constituent particles obey the same laws of physics as other particles.
The issue of how to derive Born rule probabilities in such a unitary world has been extensively discussed in the 
literature \cite{Everett57,EverettBook,Deutsch99,Saunders02,Wallace02,Wallace03} --- for thorough criticism and defense of these derivations, see \cite{SaundersBook,everett3}, and for a subsequent derivation using inflationary cosmology, see \cite{born}.
The key point of the derivations is that in unitary cosmology, a given quantum measurement tends to have multiple outcomes as illustrated in \fig{ExpandingUniverseSplitFig}, 
and that a generic rational observer can fruitfully act as if some non-unitary random process (``wavefunction collapse'') realizes only one of these outcomes at the moment of measurement, with a probabilities given by the Born rule.
This means that in the context of unitary cosmology, what is traditionally called the Copenhagen Interpretation is more aptly termed the {\it Copenhagen Approximation}: 
an observer can make the convenient approximation of pretending that the other decohered wave function branches do not exist and that wavefunction collapse does exist. In other words, the approximation is that apparent randomness is fundamental randomness.

In summary, if you are one of the many observers in \fig{ExpandingUniverseSplitFig}, you compute the density matrix $\rho$ with which to best predict your future from the 
full density matrix by performing the two complementary operations summarized in Table~1: conditioning on your knowledge (generalized 
``state preparation'') and partial tracing over the environment.\footnote{Note that the factorization of the Hilbert space into subject, object and environment subspaces
is different for different branches of the wavefunction, and that generally no global factorization exists. If you designate the spin of a particular silver atom to be your object degree of freedom in this branch of the wavefunction, then a copy of you in a branch where planet Earth (including you, your lab and said silver atom) are a light year further north will settle on a different tripartite partition into subject, object and environment degrees of freedom.
Fortunately, all observers here on Earth here in this wavefunction branch agree on essentially the same entropy for our observable universe, 
which is why we tend to get a bit sloppy and hubristically start talking about ``the'' entropy, as if there were such a thing.
}

\subsection{Outlook}


Using our tripartite decomposition formalism, we showed that because of the long-range entanglement created by cosmological inflation, the cosmic entropy decreases exponentially rather than linearly with the number of bits of information observed, so that a given observer can produce much more negentropy than her brain can store. 
Using this result, we argued that as long as inflation has occurred in a non-negligible fraction of the volume, almost all sentient observers will find themselves in a post-inflationary low-entropy Hubble volume, and we humans have no reason to be surprised that we do so as well, which solves the so-called inflationary entropy problem.
As detailed in Appendix~\ref{SnapSec}, an arguably worse problem for unitary cosmology involves gamma-ray-burst constraints on the ``Big Snap'', a fourth cosmic doomsday scenario alongside the ``Big Crunch'', ``Big Chill'' and ``Big Rip'', where an increasingly granular nature of expanding space modifies our effective laws of physics, ultimately killing us.

Our tripartite framework also  clarifies when the popular quantum gravity approximation $G_{\mu\nu}\approx 8\pi G \expec{T_{\mu\nu}}$ is valid, and how problems with recent attempts to explain dark energy as gravitational backreaction from super-horizon scale fluctuations can be understood as a failure of this approximation.
In the future, it can hopefully shed light also on other thorny issues involving quantum mechanics and macroscopic systems.


\bigskip

{\bf Acknowledgments:}
The author wishes to thank Anthony Aguirre and Ben Freivogel for helpful discussions about the degree-of-freedom problem,
Philip Helbig for helpful comments, Harold Shapiro for help proving that $S(\rho\circ\E)>S(\rho)$ 
and Mihaela Chita for encouragement to finish this paper after years of procrastination.
This work was supported by NSF grants AST-0708534,AST-090884 \& AST-1105835.
\appendix

\section{Entropy inequalities for observation and decoherence }
\label{EntropyProofSec}

\subsection{Proof that decoherence increases entropy}

The decoherence formula from Table~1 says that 
the effect of decoherence on the object  density matrix $\rho$ is
\beq{DecoSchurEq}
\rho \mapsto \rho\circ\E,
\eeq
where the matrix $\E$ is defined by $E_{ij}\equiv \langle\epsilon_j|\epsilon_i\rangle$ and the symbol 
$\circ$ denotes what mathematicians know as the {\it Schur product}.
Schur multiplying two matrices simply corresponds to multiplying their corresponding components, \ie, 
$(\rho\circ\E)_{ij}=\rho_{ij} E_{ij}$.
Because $\E$ is the matrix of inner products of all the resulting environmental states $|\epsilon_i\rangle$, it is a so-called Gramian matrix and guaranteed to be positive semidefinite (with only non-negative eigenvalues). Because $\E$ also has the property that all its diagonal elements are unity ($E_{ii}\equiv \langle\epsilon_i|\epsilon_i\rangle=1$), it is conveniently thought of as a (complex) correlation matrix.

We wish to prove that decoherence always increases the entropy 
 $S(\rho)\equiv -\tr\rho\log\rho$ of the density matrix, \ie, that
\beq{DecoEntropyTheorem}
S(\rho\circ\E)\ge S(\rho),
\eeq
for any two positive semidefinite matrices $\rho$ and $\E$ such $\tr\rho=1$ and $E_{ii}=1$, with equality only for the trivial case where $\rho\circ\E=\rho$, corresponding to the object-environment interaction having no effect.
Since I have been unable to find a proof of this in the literature, I will provide a short proof here.

A useful starting point is the Corollary J.2.a in \cite{MarshallBook} (their equation 7), which follows from a 
1985 theorem by Bapat and Sunder.
If states that 
\beq{MajorizationEq}
\lambda(\rho)\succ\lambda(\rho\circ\E),
\eeq
where $\lambda(\rho)$ denotes the vector of eigenvalues of a matrix $\rho$, arranged in decreasing order, 
and the symbol $\succ$ denotes {\it majorization}.
A vector with components $\lambda_1$, ...,$\lambda_n$ majorizes another vector with components 
$\mu_1$, ...,$\mu_n$ if they have the same sum and 
\beq{MajorizationDefEq}
\sum_{i=1}^j \lambda_i \ge \sum_{i=1}^j \mu_i\quad\hbox{for }j=1,\dots,n,
\eeq
\ie, if the partial sums of the latter never beat the former:
$\lambda_1\ge\mu_1$, 
$\lambda_1+\lambda_2\ge\mu_1+\mu_2$, {\etc} 
In other words, the eigenvalues of the density matrix before decoherence majorize the eigenvalues after decoherence.

Given any two numbers $a$ and $b$ where $a>b$, let us define a {\it Robin Hood transformation} as one that brings them closer together while keeping their sum constant: $a\mapsto a-c$ and $b\mapsto a+c$ for some constant $0<c<(a-b)/2$.
Reflecting on the definition of $\succ$ shows that majorization is a measure of how spread out a set of numbers are:
performing a Robin Hood transformation on any two elements of a vector will produce a vector that it majorizes, and the maximally egalitarian vector whose components are all equal $(\lambda_i=1/n)$ will be majorized by any other vector of the same length. Conversely, any vector that is majorized by another can be obtained from it by a sequence of Robin Hood transformations \cite{MarshallBook}.

It is easy to see that for a function $h$ that is concave (whose second derivative is everywhere negative), 
the quantify $h(a)+h(b)$ will increase whenever we perform a Robin Hood transformation on $a$ and $b$.
This implies that $\sum_{i=1}^n h(\lambda_i)$ increases for any Robin Hood transformation on any pair of elements, and when we replace the vector of $\lambda$-values by any vector that it majorizes.
However, the entropy of a matrix is exactly such a function of its eigenvalues:
\beq{ProofClincherEq}
S(\rho)\equiv-\tr\rho\log\rho = \sum_i h(\lambda_i),
\eeq
where the function $h(x)=-x\log x$ is concave.
This concludes the proof of \eq{DecoEntropyTheorem}, \ie, of the theorem that decoherence increases entropy.
By making other concave choices of $h$, we can analogously obtain other theorems about the effects of decoherence. 
For example, choosing $h(x)=-x^2$ proves that decoherence also increases the linear entropy $1-\tr\rho^2$.
Choosing $h(x)=\log x$ proves that decoherence increases the determinant of the density matrix, since 
$\log\det\rho=\sum_i\log\lambda_i$.

\subsection{Conjecture that observation reduces expected entropy}

The observation formula from Table~1
can be thought of as the quantum Bayes Theorem.
It says that observing subject state $i$ changes the object density matrix to 
\beq{Observation_rhoEq}
\rho^{(i)}_{jk} = {\rho_{jk} S_{ij} S_{ik}^*\over p_i},
\eeq
where $S_{ij}\equiv \langle s_i|\sigma_j\rangle$ and
\beq{Observation_pEq}
p_i = \sum_j \rho_{jj}  |S_{ij}|^2
\eeq
can be interpreted as the probability that the subject will perceive state $i$.
The resulting entropy $S(\rho^{(i)})$ can be both smaller and larger than the initial entropy $S(\rho)$, as simple examples show.
However, I conjecture that observation always decreases entropy {\it on average}, specifically, that
\beq{ObservationConjectureEq}
\sum_i p_i S\left(\rho^{(i)}\right) < S(\rho)
\eeq
except for the trivial case where $\rho^{(i)}=\rho$, where observation has no effect.
The corresponding result for classical physics holds, and was proven by Claude Shannon: here average entropy reduction equals the mutual information between object and subject, which cannot be negative.

For quantum mechanics, however, the situation is more subtle.
For example, for a system of two perfectly entangled qubits, the entropy of the first qubit is $S_1=1$ bit while the mutual information $I\equiv S_1+S_2-S_{12}=2$ bits, so the classical result would suggest that $S_1$ should drop to the impossible value of $-1$ bit whereas \eq{ObservationConjectureEq} shows that it drops to $0$ bits.
Although I have thus far been unable to rigorously prove \eq{ObservationConjectureEq}, I have performed extensive numerical tests with random matrices without encountering any counterexamples.

\section{The Degree-of-Freedom Problem and the Big Snap}
\label{SnapSec}

Let $N$ denote the number of degrees of freedom in a  finite comoving volume $V$ of space.
Does $N$ stay constant over time, as our universe expands? There are three logically possible answers to this questions, none of which appears problem free:
\begin{enumerate}
\itemsep0mm
\item {\bf Yes} 
 \item {\bf No}
\item {\bf $N$ is infinite},  so we don't need to give a yes or no answer.
\end{enumerate}

Option 3 has been called into doubt by quantum gravity considerations.
First, the fact that our classical notion of space appears to break down below the Planck scale $\rplanck\sim 10^{-34}$m calls into question whether $N$ can significantly exceed $V/\rplanck^3$, the volume $V$ that we are considering, measured in Planck units.
Second, some versions of the so-called holographic principle \cite{tHooft93} suggest that $N$ may be smaller still, bounded not by the
$V/\rplanck^3$ but by $V^{2/3}/\rplanck^2$, roughly the area of our volume in Planck units.
Let us therefore explore the other two options: 1 and 2. The hypothesis that degrees of freedom are neither created nor destroyed underlies not only quantum mechanics (in both its standard form and with non-unitary GRW-like modifications \cite{GRW86}), but classical mechanics as well. 
Although quantum degrees of freedom can freeze out at low temperatures, reducing the ``effective'' number, this does not change the actual number, which is simply the dimensionality of the Hilbert space.

\subsubsection{Creating degrees of freedom}

The holographic principle in its original form \cite{tHooft93} suggests option 2, changing $N$.\footnote{More recent versions of the holographic principle have focused on the entropy of 3D light-sheets rather than 3D volumes, evading the implications below\cite{Susskind95,Bousso02}.}
Let us take our comoving volume $V$ to be our current cosmological particle horizon volume, also known as our  ``observable universe", of radius $\sim 10^{26}$m, giving a holographic bound of $N\sim 10^{120}$ degrees of freedom.
This exact same comoving volume was also the horizon volume during inflation, at the specific time when the largest-scale fluctuations imaged by the WMAP-satellite \cite{wmap7} left the horizon, but then its radius was perhaps of order $10^{-28}$m, giving a holographic bound of a measly $N\sim 10^{12}$ degrees of freedom.
Since this number is ridiculously low by today's standards (I have more bits than that even on my hard disk), new degrees of freedom must have been created in the interim as per option 2.\footnote{An even more extreme example occurs if a Planck-scale region with a mere handful of degrees of freedom generates a whole new universe with say $10^{120}$ degrees of freedom via the Farhi-Guth-Guven mechanism 
\cite{FarhiGuthGuven1990}.
}
But then we totally lack a predictive theory of physics!
To remedy this, we would need a theory predicting both when and where these new degrees of freedom are created, and also what quantum states they are created with. Such a theory would also need to explain how degrees of freedom disappear when space contracts, as during black hole formation.
Although some interesting early work in this direction has been pursued (see \eg \cite{Banks10}), it appears safe to say that no complete self-consistent theory of this type has yet been proposed that purports to describe all of physical reality.

\subsubsection{The Big Snap}
This leaves option 1, constant $N$. It too has received indirect support from quantum gravity research, in this case the AdS/CFT correspondence, which suggests that quantum gravity is not merely degree-of-freedom preserving but even unitary. 
This option suffers from a different problem which I have emphasized to colleagues for some time, and which I will call {\it the Big Snap}.

If $N$ remains constant as our comoving volume expands indefinitely, then the number of degrees of freedom per unit volume drops toward zero\footnote{Some interesting models evade this conclusion by denying that the physically existing volume can ever expand indefinitely while remaining completely ``real'' in some sense. De Sitter Equilibrium cosmology \cite{Albrecht09,Albrecht11} can be given the radical interpretation that once objects leave our cosmic de Sitter horizon, they no longer have an existence independent of what remains inside our horizon, and some holographic cosmology models have related interpretations \cite{BoussoSusskind11}.} as $N/V$.
Since a rubber band consists of a finite number of atoms, it will snap if you stretch it too much. 
Similarly, if our space has a finite number of degrees of freedom $N$ and is stretched indefinitely, something bad is guaranteed to happen eventually. 

As opposed to the rubber band case, we do not know precisely what this ``Big Snap'' will be like or precisely when it will occur. However, it is instructive to consider the length scale $a\equiv (V/N)^{1/3}$: if the degrees of freedom are in some sense rather uniformly distributed throughout space, then $a$ can be thought of as the characteristic distance between degrees of freedom, and we might expect some form of spatial granularity to manifest itself on this scale. 
As the universe expands, $a$ grows by the same factor as to the cosmic scale factor, pushing this granularity to larger scales.
It is hard to imagine business as usual once $a\simgt 10^{26}$m so that the number of degrees of freedom in our Hubble volume has dropped below 1. However, it is likely that our universe will become uninhabitable long before that, perhaps when the number of degrees of freedom per atom drops below 1 ($a\simgt 1^{-10}$m, altering atomic physics) or the number of degrees of freedom per proton drops below 1 ($a\simgt 1^{-15}$m, altering nuclear physics).
This Big Snap thus plays a role similar to that of the cutoff hypersurface used to tackle the inflationary measure problem,  endowing the ``end of time'' proposal of 
\cite{BoussoEndOfTime2010} with an actual physical mechanism.

Fortunately, there are observational bounds on many types of spatial granularity from astronomical observations. 
For a simple lattice with spacing $a$, the linear dispersion relation $\omega(k)=ck$ for light gets replaced by 
$\omega(k)\propto\sin(a k)$, giving a group velocity
\beq{DispersionEq}
v = {d\omega\over d k}\propto\cos ak\approx 1-{(ak)^2\over 2} = 1-{1\over 2}\left({aE\over\hbar c}\right)^2
\eeq
as long as $a\ll k^{-1}$.
This means that if two gamma-ray photons with energies $E_1$ and $E_2$ are
emitted simultaneously a cosmological distance $c/H$ away, where $H^{-1}\sim 10^{17}$s is the Hubble time, they will reach us 
separated in time by an amount 
\beq{DeltatEq}
\Delta t \sim H^{-1}{\Delta v\over v}\sim H^{-1}\left({a\Delta E\over\hbar c}\right)^2
\eeq
if the energy difference $\Delta E\equiv |E_2-E_1|$ is of the same order as $E_1$.
Structure on a time-scale of $10^{-4}$s has been reported in the gamma-ray burst GRB 910711 \cite{Scargle97} in multiple energy bands, 
which \cite{AmelinoCamelia97} interpret as a lower bound $\Delta t\simlt  0.01\>$s for $\Delta E=200\>$keV. 
Substituting this into \eq{DeltatEq} therefore gives the constraint
\beq{aBoundEq}
a < \alimit \sim {\hbar c\over\Delta E} (H\Delta t)^{1/2}\sim 10^{-21}\,\hbox{m}. 
\eeq

If $N$ really is finite, then we can consider the fate of a hypersurface  during the early stages of inflation that is defined by $a=\astar$ for some constant $\astar$.
Each region along this hypersurface has its own built-in self-destruct mechanism, in the sense that it can only support observers like us until it has expanded by a factor $\adeath/\astar$, where $\adeath$ is the $a$-value beyond which life as we know it is impossible.
However, in the eternal inflation scenario, which has been argued to be generic \cite{Vilenkin83,Starobinsky84,LindeBook}, different regions will expand by different amounts before inflation ends, so we should expect the probability to find ourselves in a given region $\sim 10^{17}$ seconds after the end of inflation to be proportional to 
$(a/\astar)^3$ as long as $a<\adeath$, \ie, proportional to the volume of the region and hence to the number of solar systems in the region (at least for all regions that share our effective laws of physics). 
This predicts that generic observers should have $a$ drawn from the probability distribution
\beq{aProbDistEq}
f(a) = \left\{ 
\begin{tabular}{lll}
${4a^3\over\adeath^4} $	&if&$a<\adeath$,\\
$0$					&if&$a\ge\adeath$.
\end{tabular}
\right.
\eeq
The tight observational constraints in \eq{aBoundEq} are thus very surprising: even if we 
conservatively assume $\adeath=10^{-19}$m, \ie, that $a$ needs to be 10000 times smaller than a proton for us to survive, 
the probability of observing $a<\alimit$ is merely 
\beq{ImprobabilityEq}
P(a\le\alimit) = \int_0^{\alimit} f(a)da = \left({\alimit\over\adeath}\right)^4 \sim 10^{-8},
\eeq
thus ruling out this scenario at 99.999999\% confidence.
Differently put, the scenario is ruled out because it predicts that a typical (median) observer has only a couple of billion years left until the Big Snap, and has already seen the tell-tale signature of our impending doom in gamma-ray burst data.
 
This argument should obviously be taken with a grain of salt; for example, one can imagine alternative dispersion relations which weaken the bound in \eq{aBoundEq}. However, to be acceptable, 
any future theory predicting a finite unchanging number of degrees of freedom $N$ must repeat this calculation using its own formalism and successfully explain why we do not observer greater time dispersion in gamma-ray bursts.

Another important caveat is that our space is not expanding uniformly: indeed, gravitationally bound regions like our Galaxy are not expanding at all. In specific models where the degrees of freedom are localized on spatial scales smaller than galaxies, one could imagine galaxy-dwelling observers happily surviving long after intergalactic space has undergone a Big Snap, as long as deleterious effects from these faraway regions do not propagate into the galaxies. 
Note, however, that this scenario saves only the observers, not the underlying theory. Indeed, the discrepancy between theory and observation merely gets worse: repeating the above volume weighting argument now predicts that we are most likely to find ourselves alive and well in a galaxy {\it after} the Big Snap has taken place throughout most of space, so the lack of any strange observed signatures in light from distant extragalactic sources (energy-dependent arrival time differences for gamma-ray bursts, say) becomes even harder to explain.



\end{document}